\documentclass[a4paper,fleqn,usenatbib]{mnras}

\usepackage{txfonts}

\usepackage[T1]{fontenc}
\usepackage{ae,aecompl}

\usepackage{psfig}   
\usepackage{graphicx}
\usepackage{amssymb}
\usepackage{subfigure}

\title[Photoevaporation versus enrichment in the cradle of the Sun]{Photoevaporation versus enrichment in the cradle of the Sun}

\author[Patel, Polius, Ridsdill-Smith et al.]{Miti Patel$^{1,2}$\thanks{Authors made an equal contribution and each should be considered joint lead author}, Cheyenne K. M. Polius$^{1\star}$, Matthew Ridsdill-Smith$^{1,3\star}$, \newauthor Tim Lichtenberg$^4$ and Richard  J. Parker\thanks{E-mail: R.Parker@sheffield.ac.uk}\thanks{Royal Society Dorothy Hodgkin Fellow}$^1$ \vspace*{0.1cm}\\
  $^1$Department of Physics and Astronomy, The University of Sheffield, Hicks Building, Hounsfield Road, Sheffield, S3 7RH, UK \\
  $^2$School of Physics \& Astronomy, University of Leicester, University Road, Leicester, LE1 7RH, UK\\
  $^3$National Astronomical Research Institute of Thailand, 260 Moo 4, T.\,\,Donkaew, A.\,\,Maerim, Chiangmai 50180, Thailand \\
  $^4$Kapteyn Astronomical Institute, University of Groningen, PO Box 800, 9700 AV Groningen, Netherlands}
\begin{document}

\date{}
                             
\pagerange{\pageref{firstpage}--\pageref{lastpage}} \pubyear{2023}

\maketitle

\label{firstpage}

\begin{abstract}
The presence of short-lived radioisotopes (SLRs) $^{26}$Al and $^{60}$Fe in the Solar system places constraints on the initial conditions of our planetary system. Most theories posit that the origin of $^{26}$Al and $^{60}$Fe is in the interiors of massive stars, and they are either delivered directly to the protosolar disc from the winds and supernovae of the massive stars, or indirectly via a sequential star formation event. However, massive stars that produce SLRs also emit photoionising far and extreme ultraviolet radiation, which can destroy the gas component of protoplanetary discs, possibly precluding the formation of gas giant planets like Jupiter and Saturn. Here, we perfom $N$-body simulations of star-forming regions and determine whether discs that are enriched in SLRs can retain enough gas to form Jovian planets. We find that discs are enriched and survive the photoionising radiation only when the dust radius of the disc is fixed and not allowed to move inwards due to the photoevaporation, or outwards due to viscous spreading. Even in this optimal scenario, not enough discs survive until the supernovae of the massive stars and so have zero or very little enrichment in $^{60}$Fe. We therefore suggest that the delivery of SLRs to the Solar system may not come from the winds and supernovae of massive stars.
\end{abstract}


\section{Introduction}

Understanding the origin of our Solar system is one of the crucial unsolved problems in astrophysics, as constraining the initial conditions of our planetary system will also enable greater understanding of the many thousands of extrasolar planets.

It is clear that planetary systems form very early on in a star's life \citep{Haisch01,Najita2014,Richert18,Alves20,SeguraCox20},  when the star is still embedded in its natal molecular cloud in the company of tens to thousands of stellar siblings \citep{Lada03}.

Therefore, the star-forming environment likely governs the early evolution of planetary systems. Our Solar system is no exception. The decay products of short-lived radioisotopes $^{26}$Al and $^{60}$Fe in the chondritic meteorites suggest that the Sun either formed from material enriched in these isotopes \citep{Gounelle12,Gounelle15,Young2014,Boss13,Boss17}, or the isotopes were delivered to the Solar system through e.g.\,\,the supernovae and/or winds of nearby massive stars \citep{Ouellette10,Lichtenberg16b,Fatuzzo15,Fatuzzo22,Parker23a}.

However, whilst massive stars have been shown to enrich the early Solar system (and probably many other planetary systems), they have also been shown to have a very destructive influence on protoplanetary discs, as their far and extreme ultraviolet radiation acts to evaporate the gas content from protoplanetary discs \citep{Scally01,Adams04,Nicholson19a}. Photoevaporation of protoplanetary discs is a very efficient process, and occurs even in star-forming regions with relatively low stellar densities \citep{Parker21a}.

This creates an interesting tension; could the Solar system be enriched in short-lived radioactive isotopes, but retain enough of its gaseous protoplanetary disc such that the gas giant planets (Jupiter and Saturn) would still be able to form? \citet{Lichtenberg19} found that the heating from $^{26}$Al determines the long-term water budget of a planet, with Solar system-levels of $^{26}$Al resulting in liquid water oceans, and lower levels leading to ice worlds. \citet{Gilmour09} even argue that intelligent civilisations require heating from $^{26}$Al in order to form and evolve.

In this paper, we follow the enrichment of the Solar system through stellar winds and supernovae, but also track the depletion of gas in the protosolar disc due to external photoevaporation from the same massive stars that provide the enriching SLRs.

Many studies have shown that SLRs can be delivered to the protosolar disc via supernovae \citep[e.g.][]{Ouellette07,Ouellette10,Parker14a,Lichtenberg16b,Nicholson17}, and recently \citet{Parker23a} demonstrated that enrichment from the winds of massive stars can enhance (and often dominate) the delivery of SLRs to protoplanetary discs.

However, none of the above studies \citep[with the exception of][who implemented viscous spreading]{Lichtenberg16b} followed the evolution of the disc, especially under the influence of external photoevaporation, which numerous studies have shown to have a highly detrimental effect on gaseous discs in star-forming regions \citep{Scally02,Adams04,Haworth18b,Winter18b,Nicholson19a,ConchaRamirez21,Parker21a}. \citet{Zwart19} tracked the enrichment of stars from winds and supernovae, but concentrated on interactions with other low-mass stars that would sculpt the Edgeworth-Kuiper Belt and/or Oort Cloud, rather than photoevaporation of the protoplanetary discs.

The paper is organised as follows. In Section~\ref{methods} we describe the set-up of the $N$-body simulations and our post-processing analysis to calculate the evolution and enrichment of the protoplanetary discs. In Section~\ref{results} we present our results, and we provide a discussion in Section~\ref{discuss}. We conclude in Section~\ref{conclude}.

\section{Method}
\label{methods}

In this Section we describe the set-up of our numerical $N$-body simulations, and the post-processing analysis used to calculate the evolution of protoplanetary discs and the enrichment from massive star winds and supernovae.

\subsection{Star-forming regions}

Observations of young star-forming regions \citep{Cartwright04,Peretto06,Sanchez09,Hacar13,Foster15,Buckner19} show that pre-main sequence stars form in filamentary structures, resulting in a spatially and kinematically substructured distribution of the stars. Similarly, hydrodynamic simulations that follow the conversion of gas to stars also produce spatial and kinematic substructure \citep{Bate12,Dale12a}, although \citet{Parker15c} show that the relation between the gas and stars is non-trivial.

To mimic the spatially and kinematically substructured distributions in young star-forming regions, we set up our $N$-body simulations as self-similar, fractal distributions \citep{Goodwin04c}. We use the box-fractal method, which has the advantage that the amount of substructure is defined by just one number, the fractal dimension, $D$.

The spatial distribution is set up by first defining a cube with side of length $N_{\rm div} = 2$, and a `parent' star is placed at the centre of this cube. The cube is then divided into $N^3_{\rm div}$ sub-cubes and a `child' particle is placed at the centre of each sub-cube. The probability that a child particle becomes a parent particle is $N_{\rm div}^{(D-3)}$, and child particles that do not become parents are removed, along with their parent stars.

Each child's sub-cube is itself then divided into $N^3_{\rm div}$ sub-sub-cubes, and the process is repeated until there is a generation of particles with significantly more stars than required.

All of the parental generation particles are removed, so that only the final generation is retained. The distribution is then pruned so that the star particles lie within the boundary of a sphere, rather than the cube we used to create the fractal. If there are more stars remaining than the desired number of stars, $N_\star$ then stars are removed until we reach $N_\star$.

The stellar velocities follow the spatial substructure such that stars that are close together have similar velocities, whereas more distant stars can have very different velocities, similar to the observed \citet{Larson81} relations.

During the construction of the fractal, the parent particles are assigned a velocity drawn randomly from a Gaussian distribution of mean zero. The child particles inherit their parent's velocities, plus a small random offset that decreases with the number of generations from the parent particle.

The velocities are then scaled to a global virial ratio, $\alpha_{\rm vir} = T/|\Omega|$, where $T$ and $|\Omega|$ are the total kinetic and potential energies of the stars, respectively, and virial equilibrium is defined as $\alpha_{\rm vir} = 0.5$.

In all of our simulations we adopt a fractal dimension $D = 2.0$, which results in a moderately substructured distribution.%, and for similar stellar densities produces slightly lower radiation fields (we use these radiation fields to caluclate the mass lost from the protoplanetary discs, see below) in the star-forming region than less substructured, or smoother distributions \citep{Parker21a}.

Similarly, all of our simulations are set up to be subvirial, i.e.\,\,$\alpha_{\rm vir} = 0.3$, which results in a global collapse of the star-forming region, erasing the substructure and forming a smooth, centrally concentrated star cluster \citep{Parker14b}.

In our simulations, we wish to maximise the number of massive stars that could enrich protoplanetary discs with SLRs, but also produce FUV and EUV radiation fields that could cause the evaporation of gas from protoplanetary discs. Star-forming regions follow a mass distribution of the form \citep{Lada03}
\begin{equation}
  \frac{dN}{dM} \propto M_{\rm SF}^{-2},
  \label{ECMF}
\end{equation}
where $M_{\rm SF}$ is the mass of a star-forming region. Eq.~\ref{ECMF} implies that low-mass star-forming regions are much more common than high-mass star-forming regions. However, for a typical Initial Mass Function we expect there to be more massive stars when the mass of the star-forming region is higher \citep{Elmegreen06,Weidner06,Parker07,Nicholson17}. 

We choose a total number of stars, $N_\star = 1500$, such that our star-forming regions contain enough massive stars for photoevaporation and enrichment of the discs, but not so massive that the star-forming regions would be extremely rare in the Galaxy.

We draw $N_\star = 1500$ stars from a \citet{Maschberger13} IMF, which has a probability density distribution of the form
\begin{equation}
p(m) \propto \left(\frac{m}{\mu}\right)^{-\alpha}\left(1 + \left(\frac{m}{\mu}\right)^{1 - \alpha}\right)^{-\beta}.
\label{maschberger_imf}
\end{equation}
In Eq.~\ref{maschberger_imf}, $\mu = 0.2$\,M$_\odot$ is the average stellar mass, $\alpha = 2.3$ is the high mass \citet{Salpeter55} slope, and $\beta = 1.4$ is used to describe the slope of the low-mass end of the IMF. We sample the IMF in the range 0.1 -- 50\,M$_\odot$.

Our choice for the upper limit to the IMF is informed by observations of star-forming regions with a similar number of stars to those we model here, such as the ONC, where the most massive star, $\theta^1$ Ori~C, is $\sim$40\,M$_\odot$, \citep{Lehmann10}. Some authors posit that the most massive star that can form in a region is fundamentally related to the mass of the region \citep{Weidner06}, whereas others suggest the only upper limit to the mass of a star that can form is the total mass of the region itself \citep{Elmegreen06}. In the latter scenario, the most massive star can in theory be as massive as the region (although this is unlikely in random sampling), and can also be significantly lower than the most massive stellar mass from the \citet{Weidner06} relation. The most massive stars known in the Universe are $\sim$300\,M$_\odot$ \citep{Crowther10}, but we restrict our IMF to 50\,M$_\odot$ as a compromise between the constrained most massive star--region mass relation \citep{Weidner06} and the expected outcome of randomly sampling the IMF \citep{Parker07}.

We keep the number of stars $N_\star$ constant, we keep the fractal dimension $D = 2.0$ constant, and vary the initial radii of the star-forming region. We adopt radii of 1\,pc, 2.5\,pc and 5.5\,pc, which correspond to initial median stellar densities of $1000$\,M$_\odot$\,pc$^{-3}$ (``high density''), 100\,M$_\odot$\,pc$^{-3}$ (``moderate density'') and 10\,M$_\odot$\,pc$^{-3}$ (``low density''), respectively.

The star-forming regions are evolved as a pure $N$-body simulation for 10\,Myr using the \texttt{kira} integrator within the \texttt{Starlab} package \citep{Zwart99,Zwart01}. We assume the stars all form at the same time, and we do not include stellar evolution in the simulations (although we use the models for \citet{Limongi18} to determine the SLR yields, see Section~\ref{enrich_method}). For simplicity, we do not include primordial stellar binaries in the simulations, although these are likely to be ubiquitous in star-forming regions \citep{Raghavan10,Chen13,Duchene13b,Pineda15,Ward-Duong15}.

\subsection{Disc evolution}

We follow the enrichment and destruction of the protoplanetary discs via a post-processing analysis, i.e.\,\,the discs are not included in the $N$-body simulations. We use the $N$-body simulation output (every 0.01\,Myr) to track the relative positions of the massive stars compared to the low-mass (disc-hosting) stars, and then use the distances to the massive stars to determine the amount of enrichment in SLRs, and the amount of mass lost from the discs due to EUV and FUV radiation from massive stars \citep[this has a time-step of $10^{-4}$\,Myr to accruately capture the disc evolution,][]{Parker21a}.

Each star in the mass range 0.1 -- 3.0\,M$_\odot$ is assigned a disc, and we set the initial disc masses to be 10\,per cent of the host star's mass, 
 \begin{equation}
 M_{\rm disc} = 0.1\,m_\star.
 \end{equation}
 The initial disc radii are the same in each simulation, but we run different sets of the same simulation's initial conditions but with disc radii of either $r_{\rm disc} = 10$ or 100\,au.

 We first calculate how the discs respond to the FUV and EUV radiation fields from massive stars. \citet{Haworth18a} show that very little dust is lost from discs due to the FUV and EUV fields, but significant amounts of gas are destroyed, depending on the strength of the radiation field.

To estimate the FUV field incident on each low-mass star, we determine the flux of FUV radiation $F_{\rm FUV}$ using the FUV luminosity $L_{\rm FUV}$ from \citet{Armitage00}, which in turn is derived from the stellar atmosphere models of \citet{Buser92} and \citet{Schaller92}:
\begin{equation}
 F_{\rm FUV} = \frac{L_{\rm FUV}}{4\pi d^2}.
 \end{equation}
 We use the distance to each massive star, $d$, to calculate the flux and then sum the fluxes, as all simulations contain more than one massive star (defined as $>$5\,M$_\odot$ for the purposes of generating an FUV radiation field). We convert the FUV flux from cgs units into the dimensionless \citet{Habing68} unit, $G_0 = 1.8 \times 10^{-3}$\,erg\,s$^{-1}$\,cm$^{-2}$, which is the ambient background FUV flux in the interstellar medium.

 To calculate the mass lost due to the FUV radiation field, we use the \texttt{FRIED} grid from \citet{Haworth18b}, which is a grid of mass-loss rates for different combinations of stellar mass, $G_0$, disc mass, disc radius and disc surface density. For each low mass star, we use the stellar mass, disc mass, disc radius and $G_0$ field to establish via linear interpolation the most appropriate mass-loss rate due to FUV radiation, $\dot{M}_{\rm FUV}$.
 
% \begin{equation}
% F_{\rm EUV} = \frac{L_{\rm EUV}}{4\pi d^2}.
 % \end{equation}

We calculate the mass-loss due to the EUV radiation field using Eq.~\ref{euv_equation} from \citet{Johnstone98}, which depends on the distance to the massive star $d$ and the disc radius $r_{\rm disc}$:
 \begin{equation}
\dot{M}_{\rm EUV} \simeq 8 \times 10^{-12} r^{3/2}_{\rm disc}\sqrt{\frac{\Phi_i}{d^2}}\,\,{\rm M_\odot \,yr}^{-1}.
\label{euv_equation}
\end{equation}
 $\Phi_i$ is the  ionizing EUV photon luminosity from each massive star in units of $10^{49}$\,s$^{-1}$ and is dependent on the stellar mass according to the observations of \citet{Vacca96} and \citet{Sternberg03}.

 We subtract mass from the discs every $10^{-4}$\,Myr assuming the FUV and EUV-induced mass-loss rates from the \texttt{FRIED} models and Eq.~\ref{euv_equation}. Models of mass-loss from discs usually assume material is lost from the edge of the disc (where the surface density is lowest) and so we would expect the radius to decrease in this scenario.

 Previous work \citep[e.g.][]{Haworth18b,Haworth19,Parker21a} assumes that if the surface density in the inner regions of the disc (e.g. 1\,au from the star) remains constant, then the radius is reduced linearly in the presence of photoevaporative mass-loss, thus:
   \begin{equation}
\Sigma_{\rm 1\,au} = \frac{M_{\rm disc}}{2\pi r_{\rm disc} [{\rm 1\,au}]},
%\label{rescale_disc}
   \end{equation}
   where $M_{\rm disc}$ is the total disc mass, and $r_{\rm disc}$ is the disc radius, as before, and $\Sigma_{\rm 1\,au}$ is the surface density at 1\,au. We then recalculate the disc radius after mass-loss, $r_{\rm disc}(t_k)$ using 
\begin{equation}
r_{\rm disc}(t_k) = \frac{M_{\rm disc}(t_k)}{M_{\rm disc}(t_{k-1})}r_{\rm disc}(t_{k-1}).
\label{rescale_disc}
\end{equation}
The radius after mass-loss is therefore a function of radius before mass-loss, $r_{\rm disc}(t_{k-1})$ multiplied by the ratio of the disc mass after ($M_{\rm disc}(t_k)$) and before ($M_{\rm disc}(t_{k-1})$) mass-loss. The subscript $_k$ refers to the (arbitrary) time after mass-loss, and the subscript $_{k-1}$ refers to the time before mass-loss.

As well as the inward evolution caused by mass-loss, the discs also evolve due to viscous evolution, which causes the radius to increase. We implement viscous expansion in our discs by implementing the diffusion equation \citep{LyndenBell74,Pringle81} with the parameterisation given in \citet{Hartmann98} and \citet{Hartmann09}.

In this formulation, the surface density $\Sigma$ at a given radius $R$ is 
\begin{equation}
  \Sigma = 1.4 \times 10^3 \frac{{\rm e}^{-R/(R_1t_d)}}{(R/R_1)t_d^{3/2}}\left(\frac{M_{\rm disc}(0)}{\rm 0.1 M_\odot}\right)\left(\frac{R_1}{\rm 10\,au}\right)^{-2}{\rm g\,cm}^{-2},
  \label{diffusion}
\end{equation}
where $M_{\rm disc}(0)$ is the disc mass before viscous evolution and $R_1$ is a radial scaling factor, which we set as $R_1 = 10$\,au. $t_d$ is a non-dimensional time, such that at a given physical time $t$ 
\begin{equation}
t_d = 1 + \frac{t}{t_s},
\end{equation}
and the viscous timescale, $t_s$ is given by
\begin{equation}
  t_s = 8 \times 10^4\left(\frac{R_1}{\rm 10\,au}\right)\left(\frac{\alpha}{10^{-2}}\right)^{-1}\left(\frac{M_\star}{\rm 0.5\,M_\odot}\right)^{1/2}\left(\frac{T_{100\,{\rm au}}}{\rm 10\,K}\right)^{-1}{\rm yr}.
  \label{time_scale}
\end{equation}
In Eq.~\ref{time_scale}, $\alpha$ is the disc viscosity \citep{Shakura73} and $T_{\rm 100\,au}$ is the temperature of the disc at a distance of 100\,au from the host star. We assume a disc temperature profile of the form
\begin{equation}
T(R) = T_{\rm 1\,au} R^{-q},
\end{equation}
where $T_{\rm 1\,au}$ is the temperature at 1\,au from the host star and is derived from the stellar luminosity for pre-main sequence stars by \citet{Luhman03b,Luhman04a} and \citet{Kirk10}.  We adopt $q = 0.5$ and $\alpha = 0.01$ \citep{Hartmann98}. Recent observations \citep[see e.g.][and references therein]{Manara22} suggest that $\alpha << 10^{-2}$, which would result in little viscous expansion of the discs in the timeframe of the simulations. We explore the consequences of this by running sets of simulations in which we keep the dust radius fixed. 

Given the mass of the star, we calculate the temperature at 100\,au and then calculate the viscous timescale $t_s$. We then use this to calculate the surface density $\Sigma$ as a function of radius $R$. We set the truncation radius, $R_{\rm trunc}$ as the radius $R$ where the surface density falls below $10^{-6}$g\,cm$^{-2}$. After viscous expansion, the position of the truncation radius changes, and we use the truncation radius before ($R_{\rm trunc}(t_{n-1})$) and after ($R_{\rm trunc}(t_{n})$) to calculate the change to the disc radius after viscous expansion:   
 \begin{equation}
 r_{\rm disc}(t_n) = r_{\rm disc}(t_{n - 1})\frac{R_{\rm trunc}(t_n)}{R_{\rm trunc}(t_{n-1})}.
 \end{equation}
 Note that the subscripts $_n$ and $_{n-1}$ differ from the subscripts used in Eq.~\ref{rescale_disc} to denote inward evolution of the disc radius due to photoevaporation because we cannot model these two physical processes at exactly the same time. We calculate the change in disc radius due to photoevaporation \emph{before} any change in radius due to viscous evolution.

 Formally the mass-loss due to photoevaporation is the gas component of the disc, but simulations have shown that the dust radius evolves in step with the gas radius \citep{Sellek20}. To test the effect of this assumption, we will perform analysis where the dust radius is kept constant, and then where the dust radius moves with the gas radius. 

 \subsection{Disc enrichment}
 \label{enrich_method}

To estimate the amount of enrichment received by a protoplanetary disc we need to determine the amount of $^{26}$Al and $^{60}$Fe deposited from the winds and supernovae of massive stars, and how this compares to the amount of the respective stable isotopes, $^{27}$Al and $^{56}$Fe. 

 We assume the usual gas-to-dust ratio of 100:1 for Solar metallicity stars, so the total dust mass, $m_{\rm dust}$, in the disc is
\begin{equation}
  m_{\rm dust} = 0.01M_{\rm disc}.
\end{equation}
The amount of $^{26}$Al is expressed in terms of the stable version of the element, $^{27}$Al, and we determine the mass of $^{27}$Al in the disc, $m_{^{27}{\rm Al}}$, from the fraction of Al in chondrites, $f_{\rm Al, CI} = 8500 \times 10^{-6}$ \citep{Lodders03}
\begin{equation}
m_{^{27}{\rm Al}} = 8500\times10^{-6} m_{\rm dust}.
\end{equation}
Similarly, the amount of $^{60}$Fe is expressed in terms of its stable version, $^{56}$Fe, and we determine the mass of  $^{56}$Fe in the disc, $m_{^{56}{\rm Fe}}$, from the fraction of Fe in chondrites, $f_{\rm Fe, CI} = 1828 \times 10^{-4}$ \citep{Lodders03}
\begin{equation}
m_{^{56}{\rm Fe}} = 1828\times10^{-4} m_{\rm dust}.
\end{equation}

For each $<$3\,M$_\odot$ star at each snapshot in time, we calculate the geometric cross section of the disc for capturing material from a supernova, 
\begin{equation}
  \eta_{\rm SN} = \frac{\pi r_{\rm disc}^2}{4\pi d^2}{\rm cos}\, \theta,
  \label{eta_sn}
  \end{equation}
where $d$ is the distance to the massive star(s) at the instant of a supernova  and $\theta$ is the inclination of the disc. Following \citet{Lichtenberg16b}, for each star we adopt $\theta = 60^\circ$ as the likely average inclination between the disc and the ejecta for a random distribution.

The supernovae times for each massive star are calculated by summing the main, and post-main sequence timescales provided in table~5 from \citet{Limongi18}. We assume the massive stars are Solar metallicity, and are rotating at 300\,km\,s$^{-1}$ \citep{deMink13} -- see \citet{Parker23a} for a comparison with models where the massive stars are not rotating. For our adopted IMF mass range \citet{Limongi18} provide data for 13, 15, 20, 25, 40 and 60\,M$_\odot$ stars, and we perform a linear interpolation for stars whose mass lies in between the values in the data table. In the rotating models, stars with initial mass 25\,M$_\odot$ explode as supernovae after 9.9\,Myr, whereas in the non-rotating models the supernovae of 25\,M$_\odot$ stars occur at 7.7\,Myr. In the models of \citet{Limongi18}, stars more massive than 25\,M$_\odot$ collapse directly to a black hole and do not explode as supernovae.

\begin{table*}
  \caption[bf]{A summary of the different initial conditions of our simulated star-forming regions. The columns show the initial radius of the star-forming region, $r_F$, the initial median local stellar density, $\tilde{\rho}$, the disc evolution physics and the stellar evolution model. The final column indicates whether the simulation is shown in a Figure in Section~\ref{results}. }
  \begin{center}
    \begin{tabular}{|c|c|c|c|c|c|}
      \hline

$r_F$ & $\tilde{\rho}$ & $r_{{\rm disc},\,i}$ & Disc evolution & Stellar evolution & Fig. \\
      \hline
      1\,pc &  $1000$\,M$_\odot$\,pc$^{-3}$ & 100\,au & Photoevap.\,\,\& dust fixed & LC18, Rotating (300\,km\,s$^{-1}$) & \\ %7 - M\\
      1\,pc &  $1000$\,M$_\odot$\,pc$^{-3}$ & 100\,au & Photoevap.\,\,\& dust fixed & LC18, non-rotating (0\,km\,s$^{-1}$) & \\ %8 - M\\
      2.5\,pc &  $100$\,M$_\odot$\,pc$^{-3}$ & 100\,au & Photoevap.\,\,\& dust fixed & LC18, Rotating (300\,km\,s$^{-1}$) & Figs.~\ref{moderate_densities_Zratios}~\&~\ref{moderate_densities_noviscous} \\ %9 - Y\\
      2.5\,pc &  $100$\,M$_\odot$\,pc$^{-3}$ & 100\,au & Photoevap.\,\,\& dust fixed & LC18, non-rotating (0\,km\,s$^{-1}$) & \\ %10 - M\\
      \hline
      1\,pc &  $1000$\,M$_\odot$\,pc$^{-3}$ & 100\,au & Photoevap.\,\,\& dust follows gas & LC18, Rotating (300\,km\,s$^{-1}$) \\ %& 6 - Y\\
      2.5\,pc &  $100$\,M$_\odot$\,pc$^{-3}$ & 100\,au & Photoevap.\,\,\& dust follows gas & LC18, Rotating (300\,km\,s$^{-1}$) & Fig.~\ref{moderate_densities_inward_dust}\\ %4 - L\\
      \hline
      1\,pc &  $1000$\,M$_\odot$\,pc$^{-3}$ & 10\,au & Photoevap.\,\,\& viscous & LC18, Rotating (300\,km\,s$^{-1}$) & \\ %3 - L\\
      1\,pc &  $1000$\,M$_\odot$\,pc$^{-3}$ & 10\,au & Photoevap.\,\,\& viscous & LC18, non-rotating (0\,km\,s$^{-1}$) & \\ %X - L\\

      1\,pc &  $1000$\,M$_\odot$\,pc$^{-3}$ & 100\,au & Photoevap.\,\,\& viscous & LC18, Rotating (300\,km\,s$^{-1}$) & \\ %5 - N\\
      1\,pc &  $1000$\,M$_\odot$\,pc$^{-3}$ & 100\,au & Photoevap.\,\,\& viscous & LC18, Non-rotating (0\,km\,s$^{-1}$) & \\ %X - N\\
            2.5\,pc &  $100$\,M$_\odot$\,pc$^{-3}$ & 100\,au & Photoevap.\,\,\& viscous & LC18, Rotating (300\,km\,s$^{-1}$) & \\ %1 - N\\ % error with previous Fig?
      2.5\,pc &  $100$\,M$_\odot$\,pc$^{-3}$ & 10\,au & Photoevap.\,\,\& viscous & LC18, Rotating (300\,km\,s$^{-1}$) & Fig.~\ref{moderate_densities_10au}\\ %2 - L\\
      2.5\,pc &  $100$\,M$_\odot$\,pc$^{-3}$ & 10\,au & Photoevap.\,\,\& viscous & LC18, Non-rotating (0\,km\,s$^{-1}$) & \\ %X - L\\

      5.5\,pc &  $10$\,M$_\odot$\,pc$^{-3}$ & 100\,au & Photoevap.\,\,\& viscous & LC18, Rotating (300\,km\,s$^{-1}$) & \\ %11 - N\\

%1\,pc & $1000$\,M$_\odot$\,pc$^{-3}$ & 2.0 & 0.3 & M13, stochastic \\
%\hline

      \hline
    \end{tabular}
  \end{center}
  \label{simulations}
\end{table*}

  Because we adopt the rotating models (in line with observations of massive stars), the majority of the $^{26}$Al originates in the winds of the massive stars \emph{before} they explode as supernovae. In their tables 8~and~9, \citet{Limongi18} provide the total yields for different isotopes (table~8) and then the yields solely from the winds of the massive stars (table~9). We use this information to determine the yield from supernovae only, and the yield from the wind over the lifetime of the massive star(s).

We assume that the isotope production in the wind is linear with time, and we use this to obtain a yield per time, which we then use to calculate the amount of $^{26}$Al and $^{60}$Fe produced by each massive star at a given time  in the simulation. In reality, the mass-loss from the stars is unlikely to be linear, and we are probably overestimating the production of $^{26}$Al at early times. Similarly, we do not account for the radioactive decay of the $^{26}$Al  and therefore our results can be thought of as an upper limit to the amount of enrichment that can be obtained from the massive stars' winds. 

The cross section for capture of wind material is calculated as the volume of material swept out by a low-mass star as it traverses a distance $\Delta r_\star$ through a wind bubble of radius $r_{\rm bub}$
\begin{equation}
  \eta_{\rm wind} = \frac{3}{4}\frac{\pi r_{\rm disc}^2\Delta r_\star}{\pi r_{\rm bub}^3}.
  \label{eta_wind}
\end{equation}
We implement two different regimes for the density of the wind bubbles. First, we assume a very compact bubble around each massive star with a radius $r_{\rm bub} = 0.1$\,pc. Secondly, we assume the bubble(s) disperse rapidly, and the bubble has a radius $r_{\rm bub} = 2r_{1/2}$, where $r_{1/2}$ is the half-mass radius of the star-forming region (i.e.\,\,the radius within which half the total stellar mass in the region is enclosed).

We use the total mass of $^{26}$Al and $^{60}$Fe captured/swept up by each low mass star and divide this by the mass in stable isotopes, $m_{^{27}{\rm Al}}$ and  $m_{^{56}{\rm Fe}}$, to determine the $^{26}$Al and $^{60}$Fe ratios:
\begin{equation}
Z_{\rm Al} = \frac{m_{^{26}{\rm Al}}}{m_{^{27}{\rm Al}}},
\end{equation}
and
\begin{equation}
Z_{\rm Fe} = \frac{m_{^{60}{\rm Fe}}}{m_{^{56}{\rm Fe}}}.
\end{equation}
For each low-mass star, we calculate $Z_{\rm Al}$ three times; for supernovae only, for supernovae and local winds ($r_{\rm bub} = 0.1$\,pc), and then finally for supernovae and dispersed winds  ($r_{\rm bub} = 2r_{1/2}$). The contribution of $^{60}$Fe comes from supernovae only.

\citet{Limongi18} provide different yields of $^{26}$Al and $^{60}$Fe depending on whether the massive stars are rotating at 0, 150 or 300\,km\,s$^{-1}$. For the non-rotating models (0\,km\,s$^{-1}$) massive stars explode as supernovae earlier, but less material is emitted in their winds. In contrast, the fast rotating stars have longer lifetimes and explode as supernovae later, but more material is emitted in their winds.\\

 A summary of the different simulations (and disc physics) is given in Table~\ref{simulations}.

\section{Results}
\label{results}

In this Section we will present three sets of simulations, each with a different treatment of the protoplanetary disc physics. We will then discuss the effects of varying other initial conditions, such as the stellar density and the stellar evolution models used to calculate the amount of $^{26}$Al and $^{60}$Fe delivered to the protoplanetary discs.

\subsection{Photoevaporation with fixed dust radius, $r_{\rm disc,i} = 100$\,au}
%2.5\,pc, 100\,au, photoevap \& dust fixed, LC18r3, Fig. 9}

The models in \citet{Parker23a} assumed the disc radius remained constant throughout the duration of the simulation. In this first set of models from our initial conditions, we adjust the gas radius of the disc according to Eqn.~\ref{rescale_disc}, and we also record the reduction in the disc's (gas) mass. However, the dust radius (and mass) is assumed to remain constant, and we therefore calculate the $Z_{\rm Al} = ^{26}$Al/$^{27}$Al and  $Z_{\rm Fe} = ^{60}$Fe/$^{56}$Fe ratios assuming a dust radius $r_{\rm disc} = 100$\,au.

\begin{figure*}
  \begin{center}
\setlength{\subfigcapskip}{10pt}
\hspace*{0.2cm}\subfigure[]{\label{moderate_densities_Zratios-a}\rotatebox{270}{\includegraphics[scale=0.33]{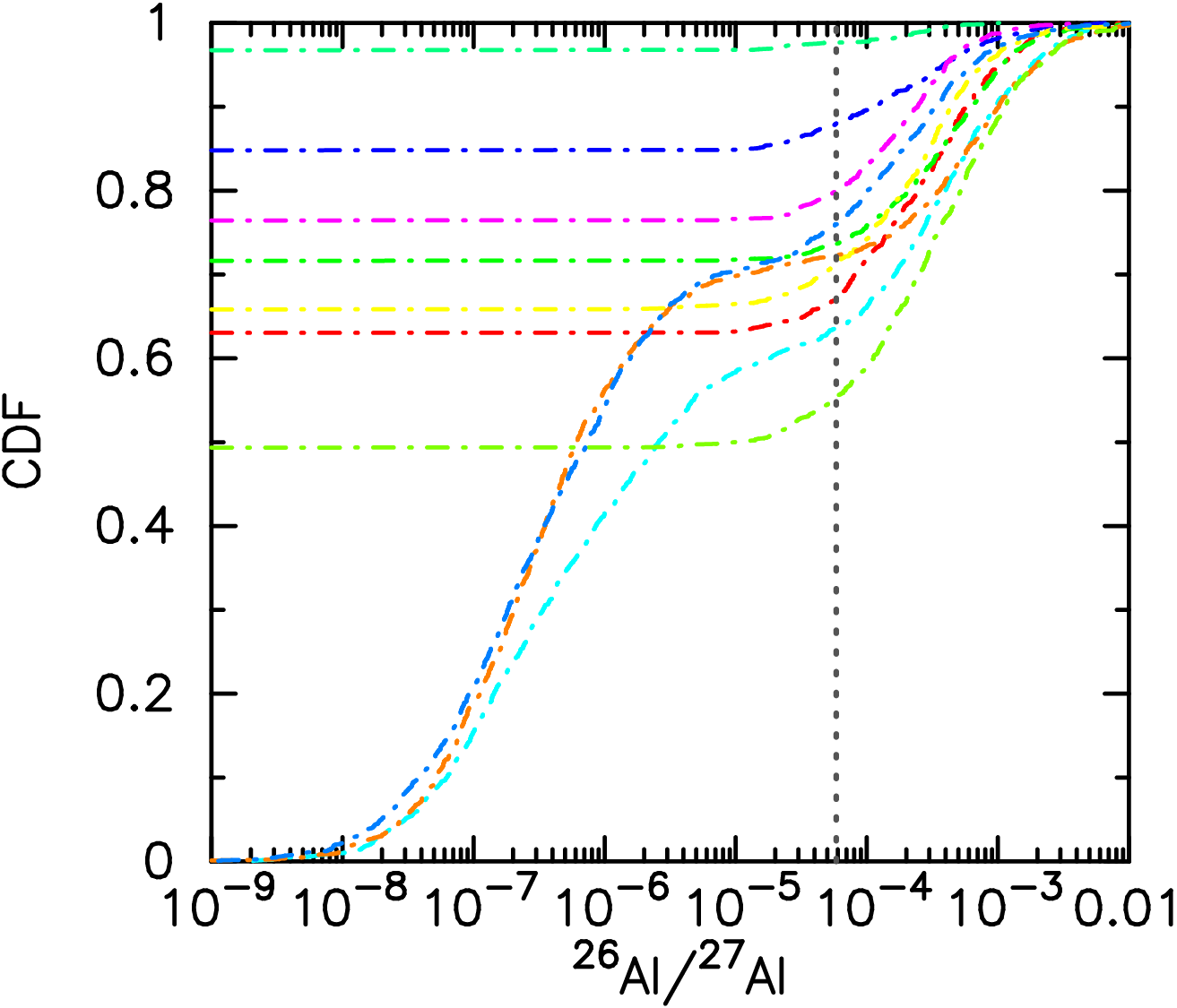}}}
\hspace*{0.2cm}\subfigure[]{\label{moderate_densities_Zratios-b}\rotatebox{270}{\includegraphics[scale=0.33]{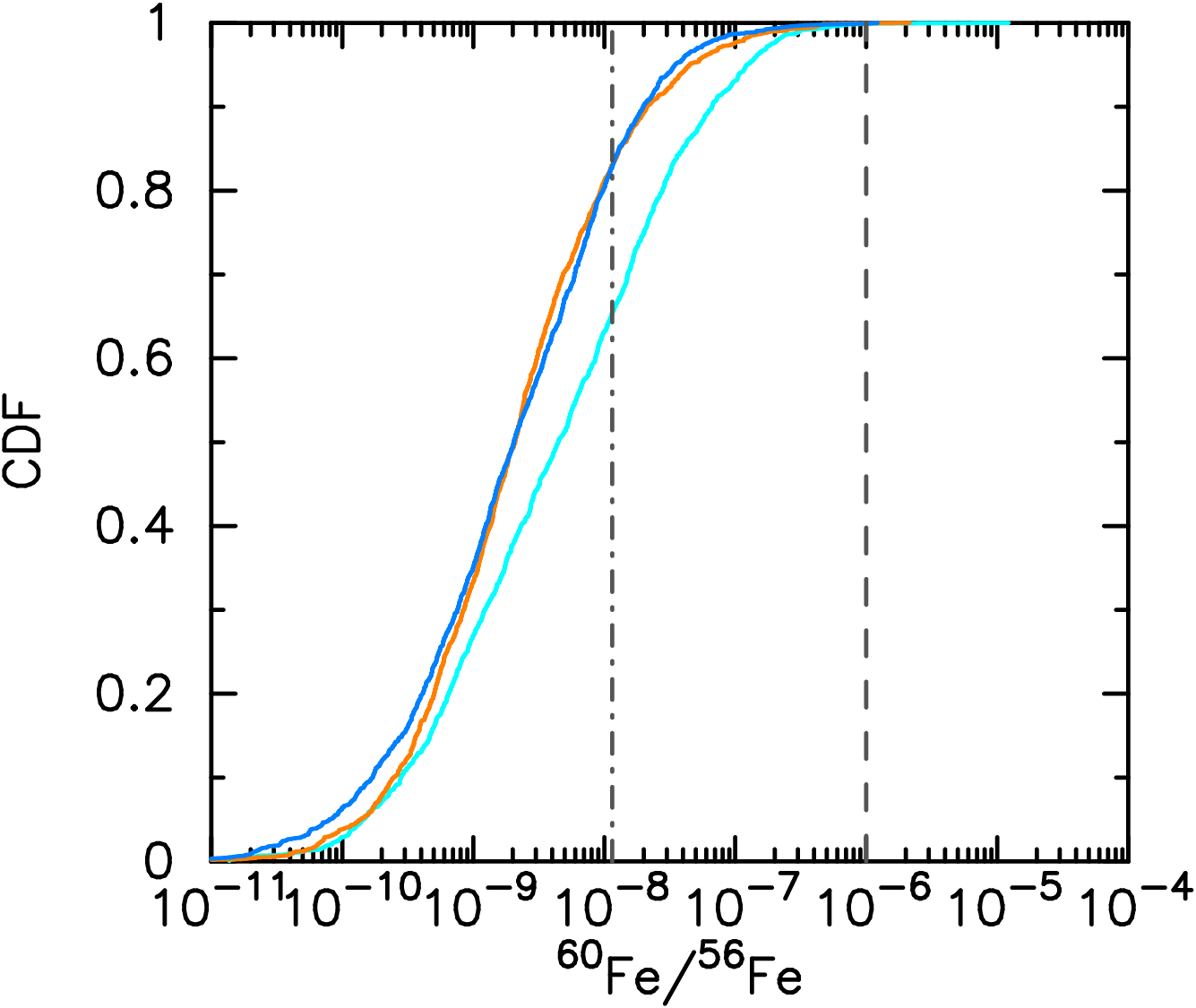}}}
\caption[bf]{The abundances of $^{26}$Al (panel a) and $^{60}$Fe (panel b) after 10\,Myr in our moderately dense ($\tilde{\rho} \sim 100$\,M$_\odot$\,pc$^{-3}$) simulations where the disc loses gas mass, and the gas radius moves inwards, but the dust radius is fixed to the initial value ($r_{\rm disc} = 100$\,au). Each line represents the distribution of low-mass ($\leq$3\,M$_\odot$) disc-hosting stars in each separate simulated star-forming region. Only three simulations contain supernovae that explode before 10\,Myr (the orange, dark blue and cyan lines in both panels). The $^{26}$Al enrichment shown in panel (a) is from supernovae and localised stellar winds. In seven simulations, a significant fraction (shown where the coloured lines meet the y-axis in panel (a)) do not experience any enrichment.  The vertical dotted line in panel (a) shows the measured $^{26}$Al abundance in the Solar system \cite{Thrane06}. The vertical dot-dashed line in panel (b) shows the (lower) estimate for the measured $^{60}$Fe abundance in the Solar system \citep{Tang12}, whereas the vertical dashed line shows the (higher) estimate for the measured $^{60}$Fe abundance in the Solar system \citep{Mishra16}.}
%Simulation 2, 10\% disc mass, 2D, distance to most massive star. 
\label{moderate_densities_Zratios}
  \end{center}
\end{figure*}

We first show the expected enrichment distributions for these simulations in Fig.~\ref{moderate_densities_Zratios}. In panel (a) we show the $Z_{\rm Al}$ ratios in these simulations, where we assume the enrichment comes from both supernovae and stellar winds, where the winds are restricted to local ($<0.1$\,pc) bubbles around the massive star(s). Each line is the $Z_{\rm Al}$ distribution in a separate run of the same simulation initial conditions, and comprises all of the low-mass ($\leq3$\,M$_\odot$) stars in each simulation. Due to the stochastic sampling of the IMF, the total number of low-mass (disc-hosting) stars varies between simulations, but is usually higher than 1450 out of the total number of 1500 stars in each simulation.   The vertical dotted line is the measured value for the Solar system \citep[$Z_{\rm Al,SS} = 5.85 \times 10^{-5}$,][]{Thrane06}

Supernovae explode before 10\,Myr in only three out of ten simulations. This may seem low, but the \citet{Limongi18} models we adopt assume stellar rotation and that massive stars $>$25\,M$_\odot$ collapse directly to form a  black hole. The stellar rotation delays the supernovae, and stars less massive than 25\,M$_\odot$ do not explode until very late on in the simulations.

In panel (b) we show the enrichment distributions for $Z_{\rm Fe}$, where the enrichment comes from supernovae only. Here, almost no systems attain the levels of enrichment commensurate with the high estimate for the Solar system \citep[$10^{-6}$,][the right vertical dashed line in Fig.~\ref{moderate_densities_Zratios-b}]{Mishra16,Cook2021}; and around 25\,per cent of systems experience enrichment  commensurate with the low estimate for the Solar system \citep[$10^{-8}$,][the left vertical dashed line in Fig.~\ref{moderate_densities_Zratios-b}]{Tang12,Trappitsch2018}.

We now determine that -- for the systems that are enriched -- how many of their discs survive the photoevaporation from the same massive stars that are enriching the discs.

The star-forming regions have moderate initial stellar densities ($\tilde{\rho} \sim 100$\,M$_\odot$\,pc$^{-3}$) and radiation fields of $10^3 - 10^4G_0$ \citep{Parker21a}. The fraction of discs that lose all of their gas in these simulations is $\sim$60\,per cent after 2.5\,Myr and 90\,per cent after 10\,Myr.

\begin{figure*}
  \begin{center}
\setlength{\subfigcapskip}{10pt}

\hspace*{-1.0cm}\subfigure[]{\label{moderate_densities_noviscous-a}\rotatebox{270}{\includegraphics[scale=0.23]{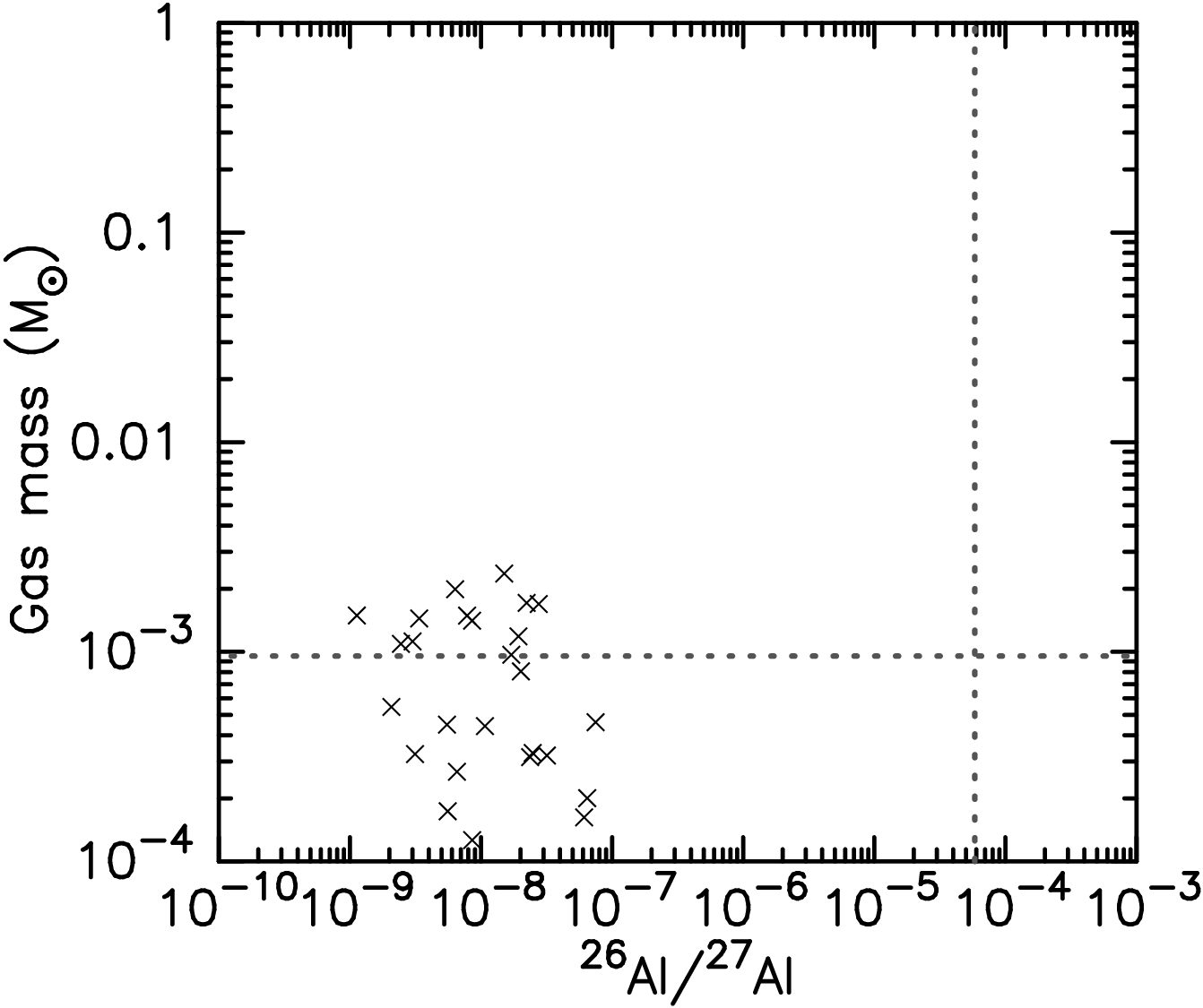}}}
\hspace*{0.2cm}\subfigure[]{\label{moderate_densities_noviscous-b}\rotatebox{270}{\includegraphics[scale=0.23]{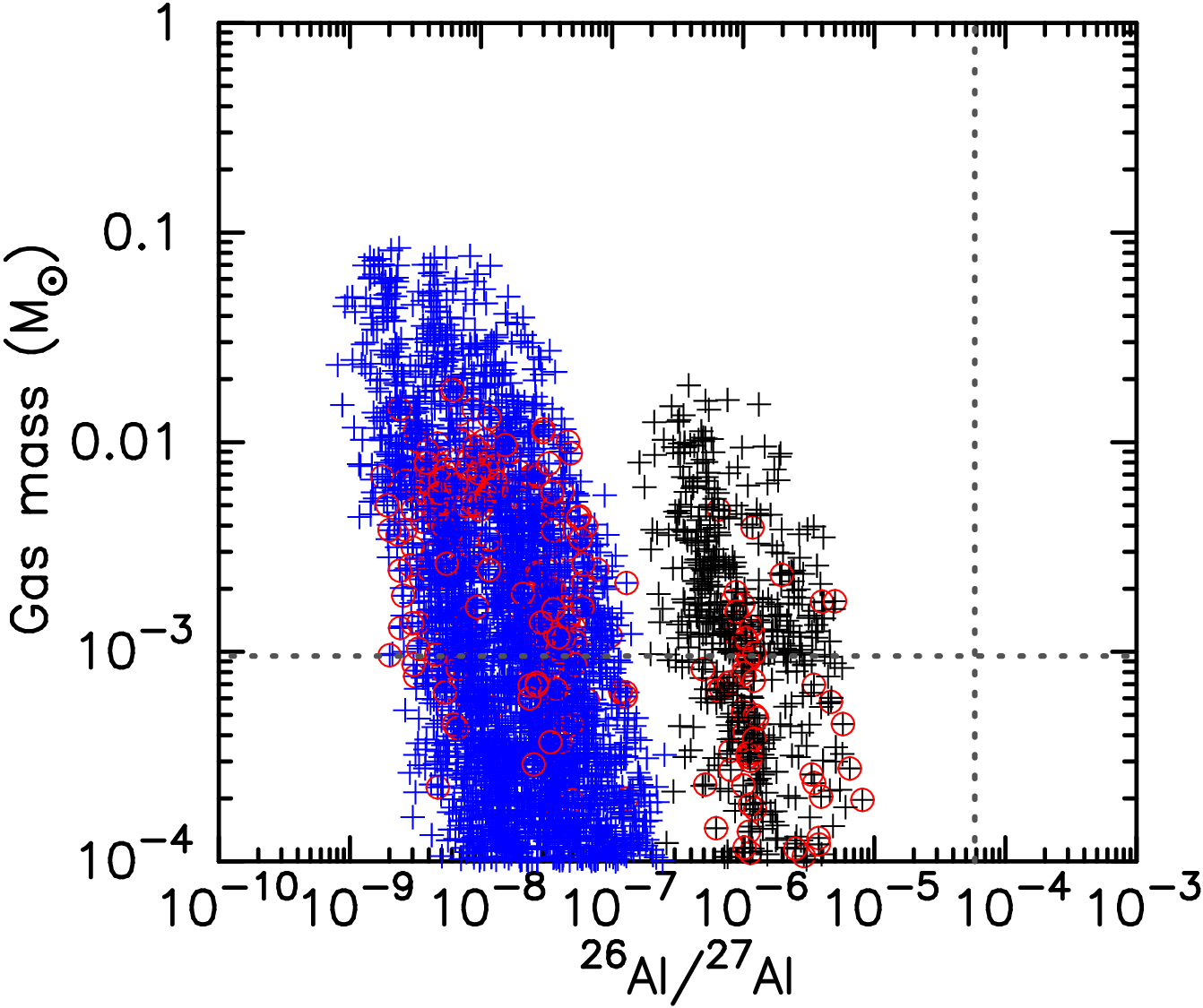}}}
\hspace*{0.2cm}\subfigure[]{\label{moderate_densities_noviscous-c}\rotatebox{270}{\includegraphics[scale=0.23]{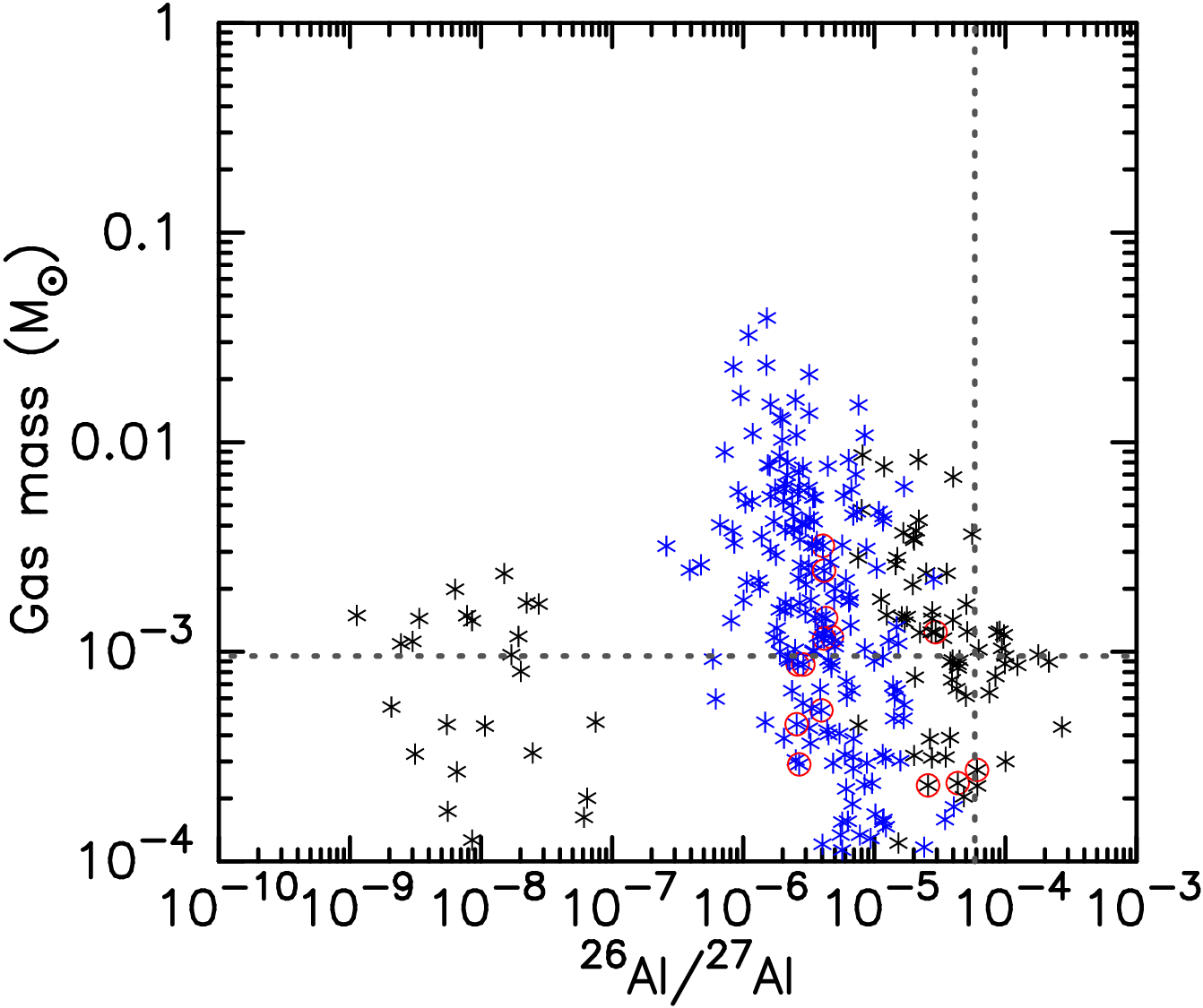}}}
\hspace*{-1.0cm}\subfigure[]{\label{moderate_densities_noviscous-d}\rotatebox{270}{\includegraphics[scale=0.23]{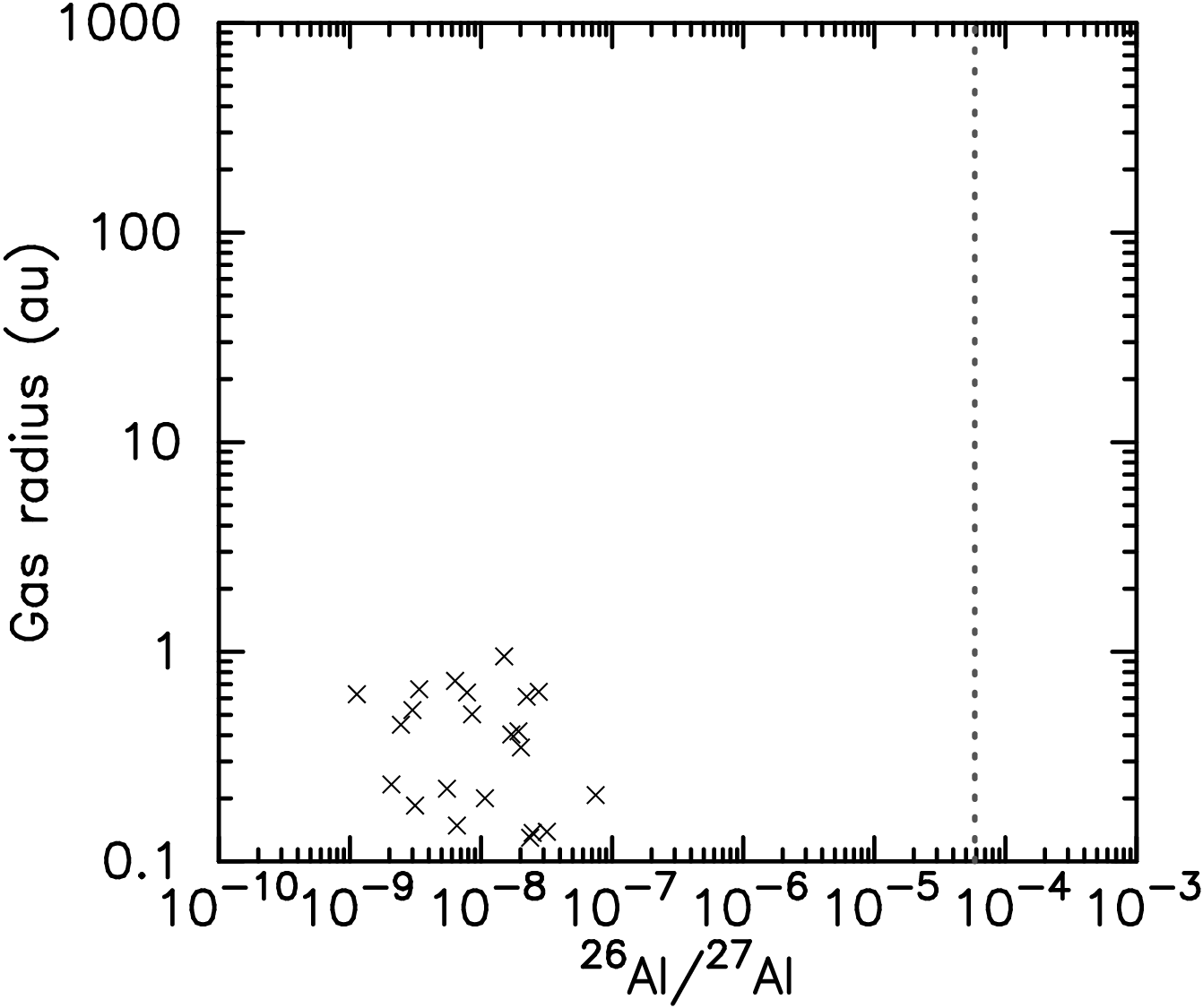}}}
\hspace*{0.2cm}\subfigure[]{\label{moderate_densities_noviscous-e}\rotatebox{270}{\includegraphics[scale=0.23]{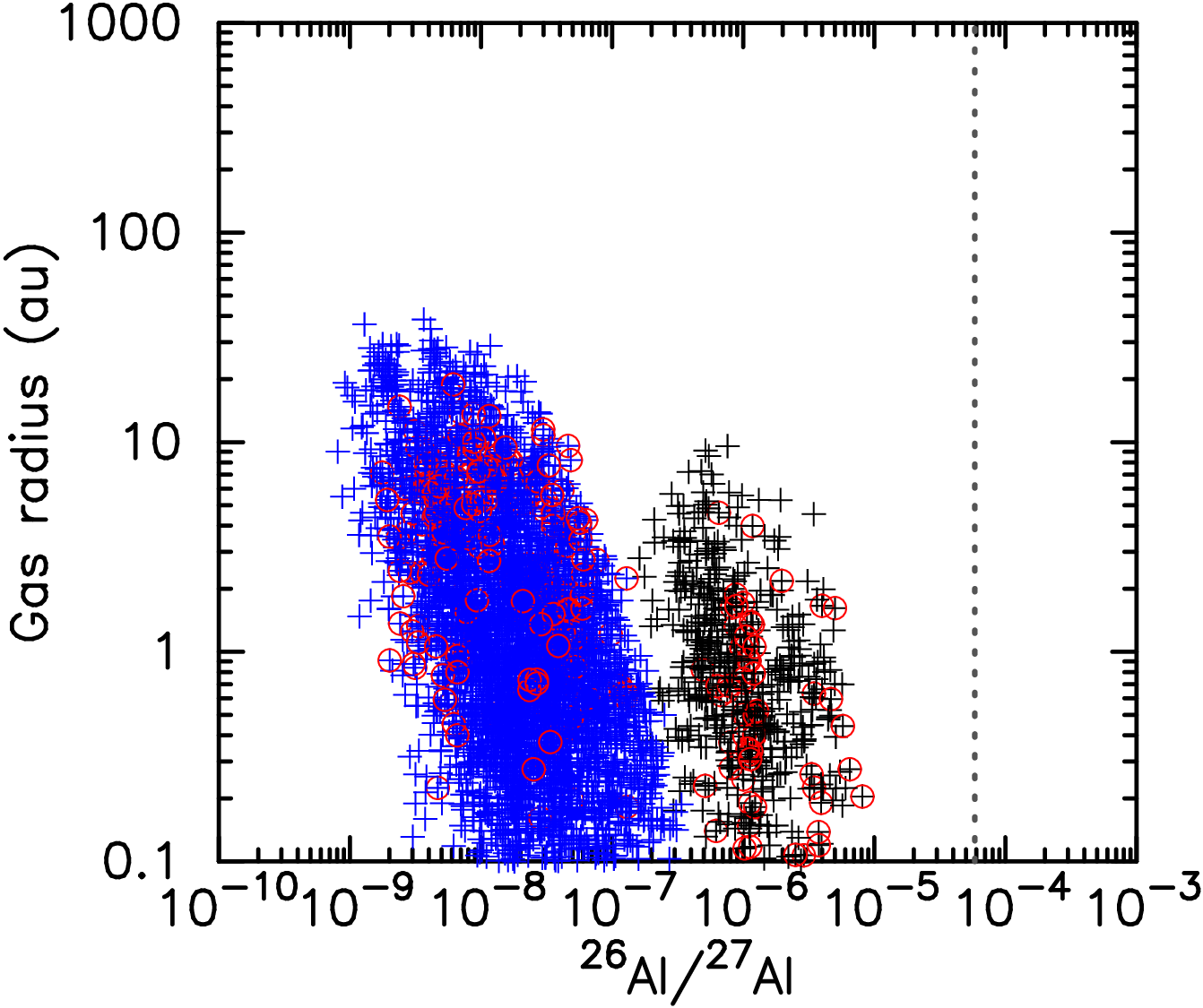}}}
\hspace*{0.2cm}\subfigure[]{\label{moderate_densities_noviscous-f}\rotatebox{270}{\includegraphics[scale=0.23]{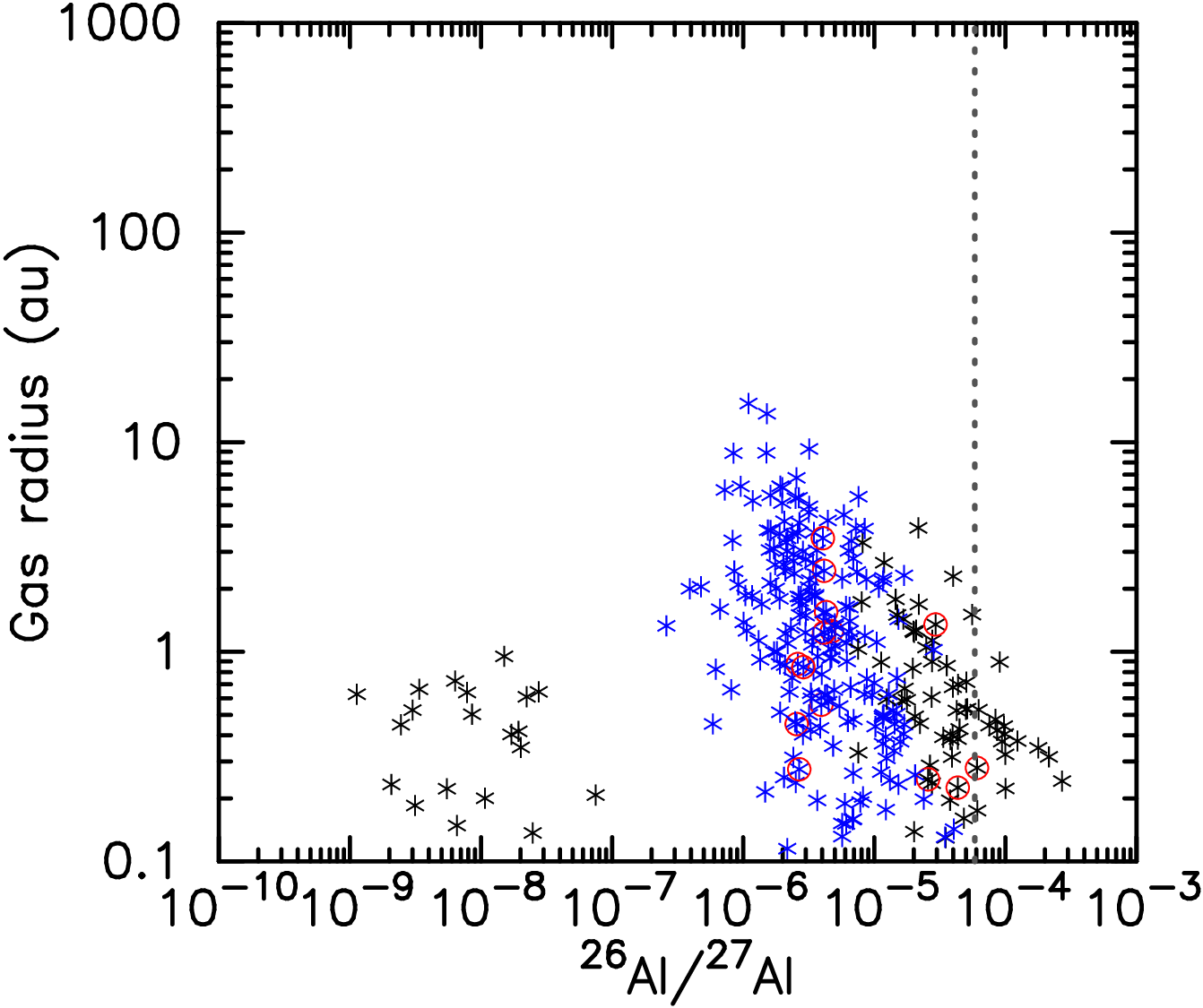}}}
\hspace*{-1.0cm}\subfigure[]{\label{moderate_densities_noviscous-g}\rotatebox{270}{\includegraphics[scale=0.23]{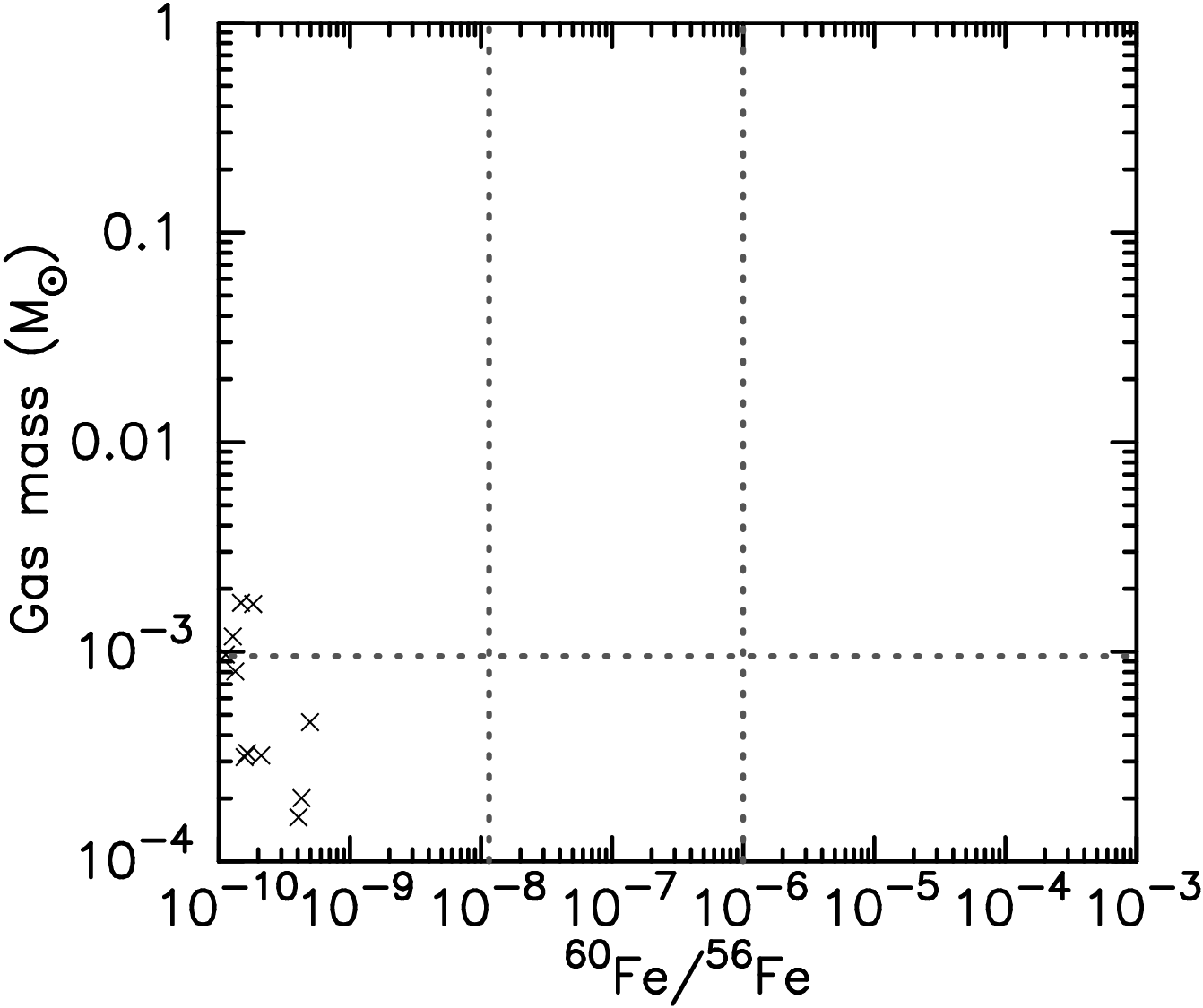}}}
\hspace*{0.2cm}\subfigure[]{\label{moderate_densities_noviscous-h}\rotatebox{270}{\includegraphics[scale=0.23]{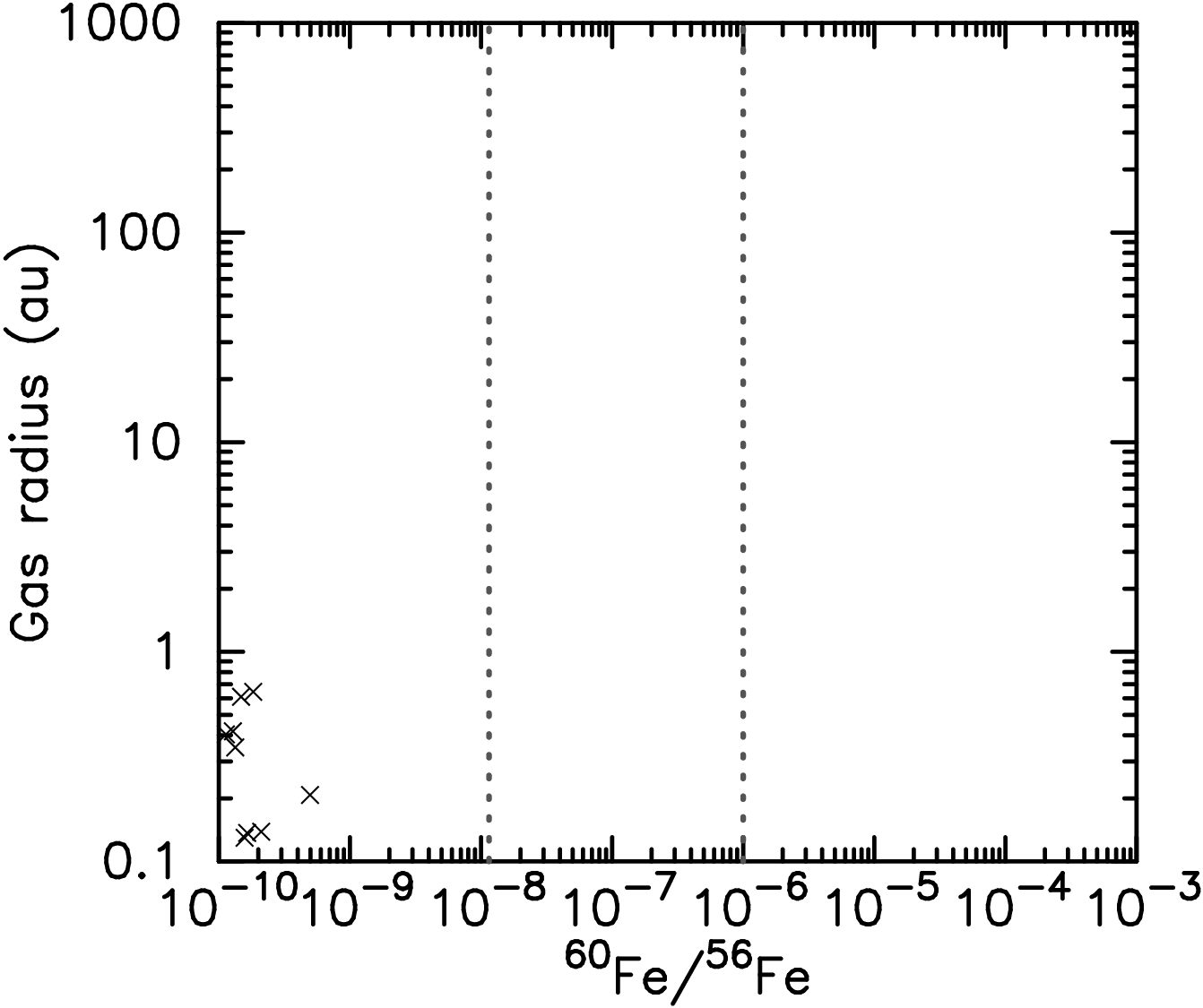}}}
\caption[bf]{Properties of discs as a function of the amount of enrichment in a moderately dense star-forming region ($\tilde{\rho} \sim 100$\,M$_\odot$\,pc$^{-3})$ where the discs all have initial radii $r_{\rm disc,i} = 100$\,au. The disc gas radii are evolved inwards due to photoevaporation (and the discs lose gas-mass), but the dust radii (and masses) remain constant. Panels (a)--(c) show the gas mass as a function of the amount of $^{26}$Al enrichment (expressed as the $^{26}$Al/$^{27}$Al ratio). Panel (a) shows enrichment from supernovae only, panel (b) shows enrichment from supernovae and dispersed wind ejecta and panel (c) shows enrichment from supernovae and localised wind ejecta. Panels (d)--(f) show the gas radii as a function of the amount of $^{26}$Al enrichment, where panel (d) shows enrichment from supernovae only, panel (e) shows enrichment from supernovae and dispersed wind ejecta and panel (f) shows enrichment from supernovae and localised wind ejecta. Panel (g) shows the disc gas mass as a function of the $^{60}$Fe enrichment and panel (h) shows the disc radius as a function of the $^{60}$Fe enrichment. Vertical dotted lines in panels (a)--(f) show the measured $^{26}$Al abundance in the Solar system \cite{Thrane06}. The vertical dotted lines in panels (g) and (h) show the range of quoted values for the measured $^{60}$Fe abundance in the Solar system \citep{Tang12,Mishra16}. The horizontal lines in panels (a)--(c) and (g) indicate 1\,M$_{\rm Jup}$, which we take to be the minimum mass required to form Jupiter and Saturn in our Solar system. In all panels, black points indicate the values at 10\,Myr, whereas the blue points indicate the values much earlier, at 2.5\,Myr. Stars that are roughly Solar-mass (0.5 - 1.5\,M$_\odot$) are highlighted by a red circle. }
\label{moderate_densities_noviscous}
  \end{center}
\end{figure*}

In Fig.~\ref{moderate_densities_noviscous} we show the disc gas properties (radius and mass) against the Al and Fe abundances. In all panels we show the values at 2.5\,Myr by the blue points, and at 10\,Myr by the black points. Panels (a)--(c) show the disc gas mass as a function of the $^{26}$Al abundance. Panel (a) shows the abundances from supernovae only, and therefore there are no data at 2.5\,Myr because the massive stars have yet to explode.

Panels (b) and (c) show the gas mass as a function of abundance from both supernovae and stellar winds, either assuming the winds are dispersed in the star-forming regions (panel b) or concentrated in local bubbles (panel c). In both cases, there are more data points at 2.5\,Myr because fewer discs have been destroyed by the EUV/FUV fields.

Interestingly, the enrichment from supernovae alone (panel a), and supernovae and dispersed winds (panel b) does not reach the levels measured in the Solar system ($Z_{\rm Al} \leq 10^{-5}$, compared to the Solar system value of $Z_{\rm Al, SS} = 5.85 \times 10^{-5}$).

However, for winds entrained in localised ($<0.1$\,pc) bubbles, several stars per simulation reach Solar system-like abundances. Note that there are no systems with low $^{26}$Al/$^{27}$Al ratios after 2.5\,Myr in the simulations where the winds are entrained in localised bubbles (the blue points in Fig.~\ref{moderate_densities_noviscous-c}~and~\ref{moderate_densities_noviscous-f}); this is because the yields from these localised bubbles are much higher than when the winds are dispersed thoughout the star-forming region. The handful of points at 10\,Myr with low $^{26}$Al/$^{27}$Al ratios are stars that are enriched by supernovae only, and previously did not encounter the small-volume wind bubbles.   Furthermore, around half of these objects still retain enough gas ($>$1\,M$_{\rm Jup}$) to form gas giant planets (systems lying above the horizontal line in panels (a)--(c) and on or around the vertical line). A tiny subset of these systems are Solar-mass stars (indicated by the red circles).

Similarly, for the gas radii (Figs.~\ref{moderate_densities_noviscous-d}--\ref{moderate_densities_noviscous-f}), around a third of the systems that are enriched to Solar system levels retain reasonably large ($>1$\,au) disc radii.

In these simulations, the enrichment in $^{60}$Fe comes from supernovae alone and the gas mass-enrichment plot is shown in Fig.~\ref{moderate_densities_noviscous-g} and the gas radii-enrichment plot is shown in Fig.~\ref{moderate_densities_noviscous-h}. The very few surviving gas-rich discs do not attain enrichment levels even at the lower-end of the measured values for our Solar system  \citep[][the lefthand vertical line in panels (g) and (h)]{Tang12}.

\subsection{Photoevaporation where dust radius follows gas radius inwards, $r_{\rm disc,i} = 100$\,au}
%2.5\,pc, 100\,au, photoevap \& dust follows gas, LC18r3, Fig. 4}

We now run the same set of simulations (moderate stellar density -- $\tilde{\rho} \sim 100$\,M$_\odot$\,pc$^{-3}$), initial disc radii $r_{\rm disc,i} = 100$\,au with external photoevaporation, but this time the dust radius follows the gas radius \citep[e.g.][]{Sellek20}. For example, when we adjust the radius due to photoevaporative mass-loss using Eq.~\ref{rescale_disc}, we also reduce the radius of the disc when calculating the cross sections for sweeping up supernovae and wind ejecta (Eqs.~\ref{eta_sn}~and~\ref{eta_wind}).

\begin{figure*}
  \begin{center}
\setlength{\subfigcapskip}{10pt}
\hspace*{0.2cm}\subfigure[]{\label{moderate_densities_inward_dust-b}\rotatebox{270}{\includegraphics[scale=0.33]{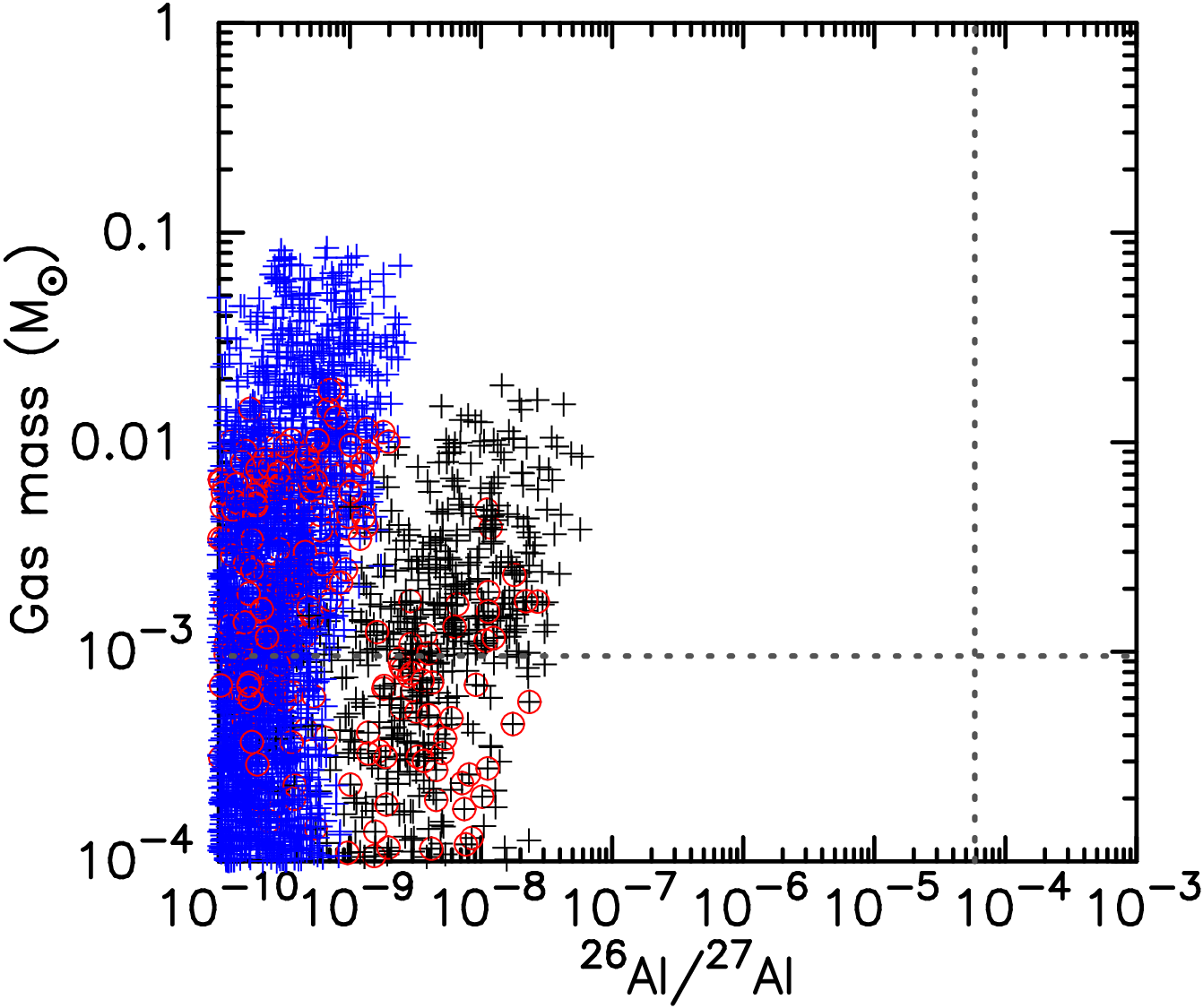}}}
\hspace*{0.2cm}\subfigure[]{\label{moderate_densities_inward_dust-c}\rotatebox{270}{\includegraphics[scale=0.33]{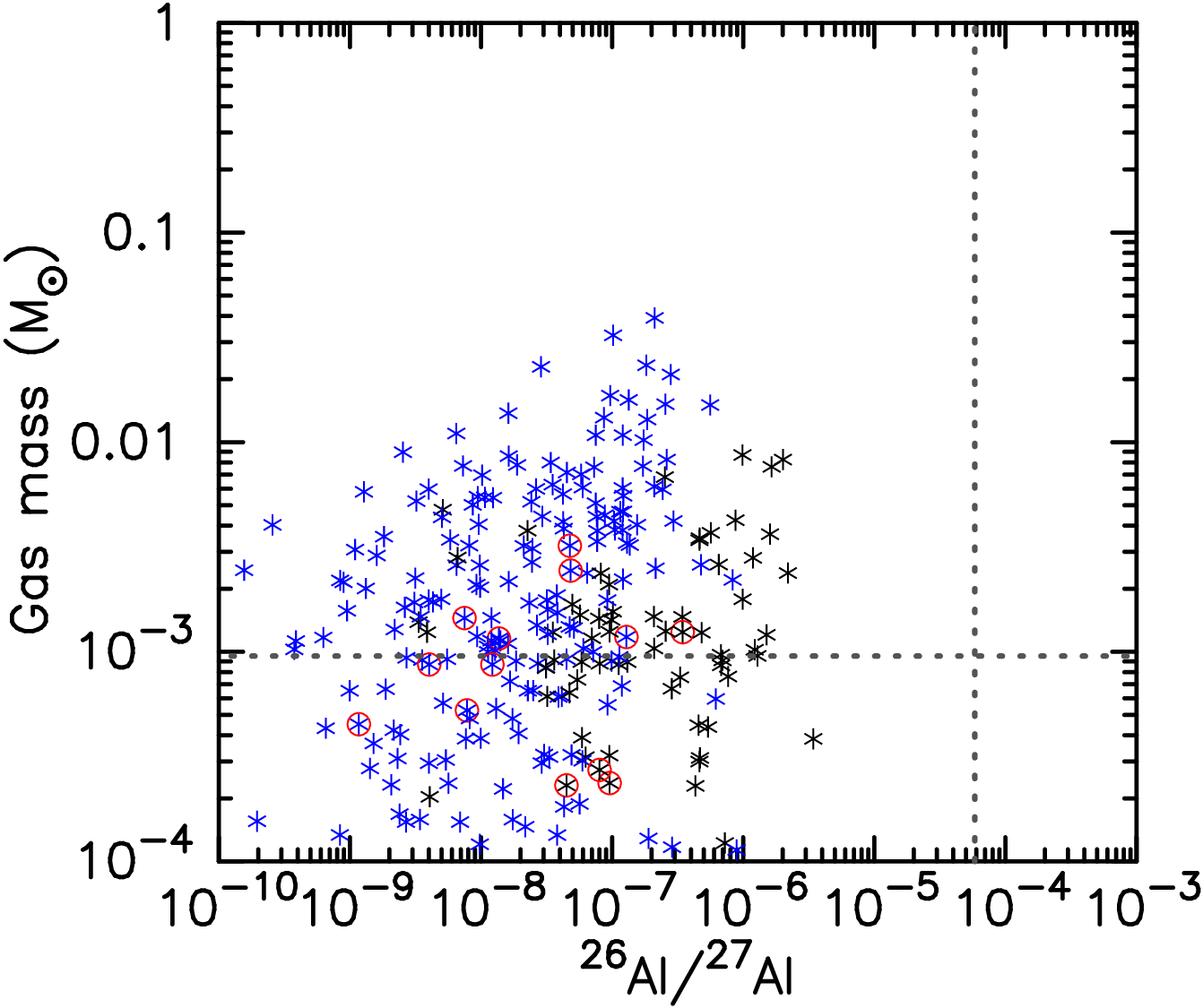}}}
\hspace*{0.2cm}\subfigure[]{\label{moderate_densities_inward_dust-e}\rotatebox{270}{\includegraphics[scale=0.33]{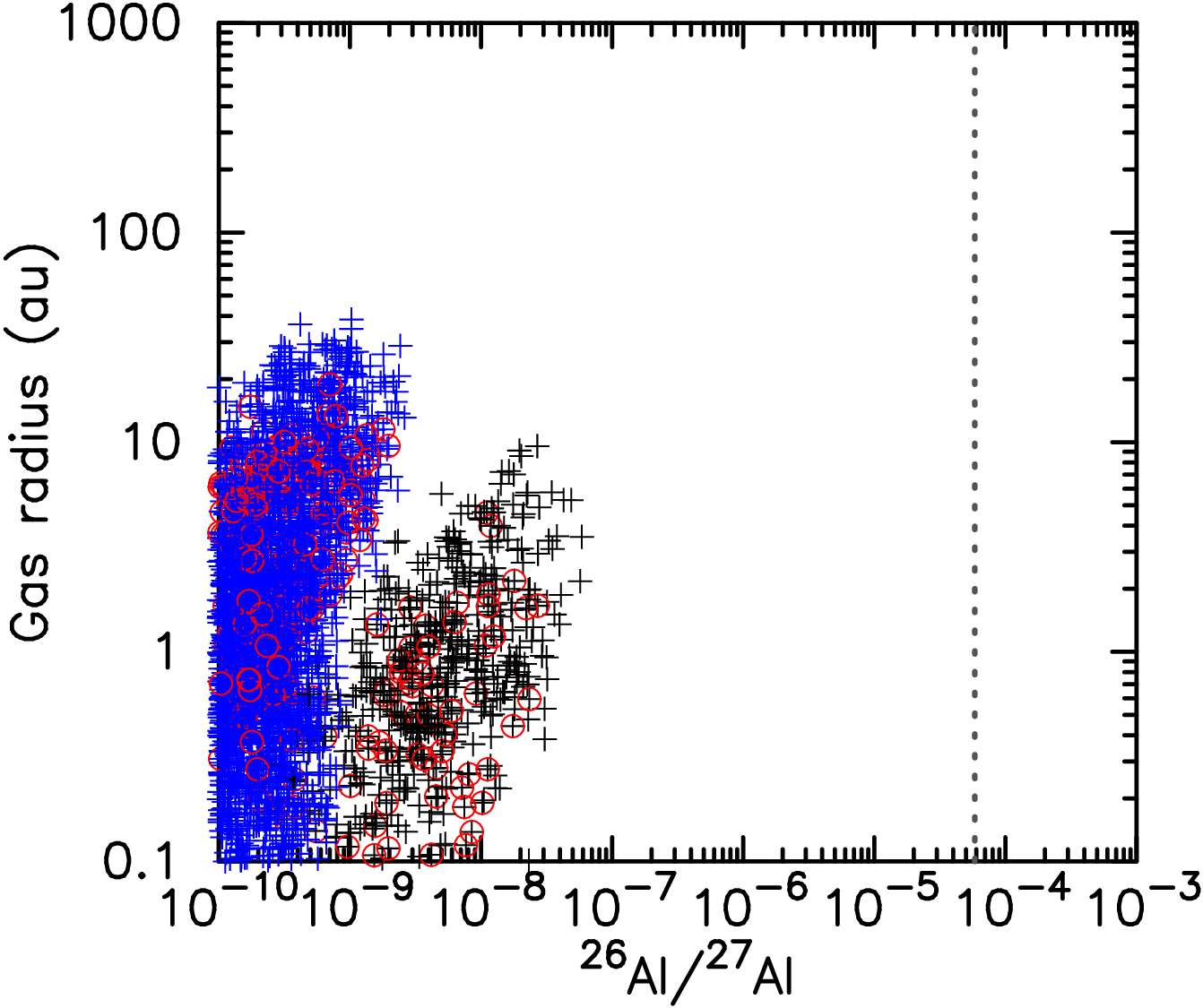}}}
\hspace*{0.2cm}\subfigure[]{\label{moderate_densities_inward_dust-f}\rotatebox{270}{\includegraphics[scale=0.33]{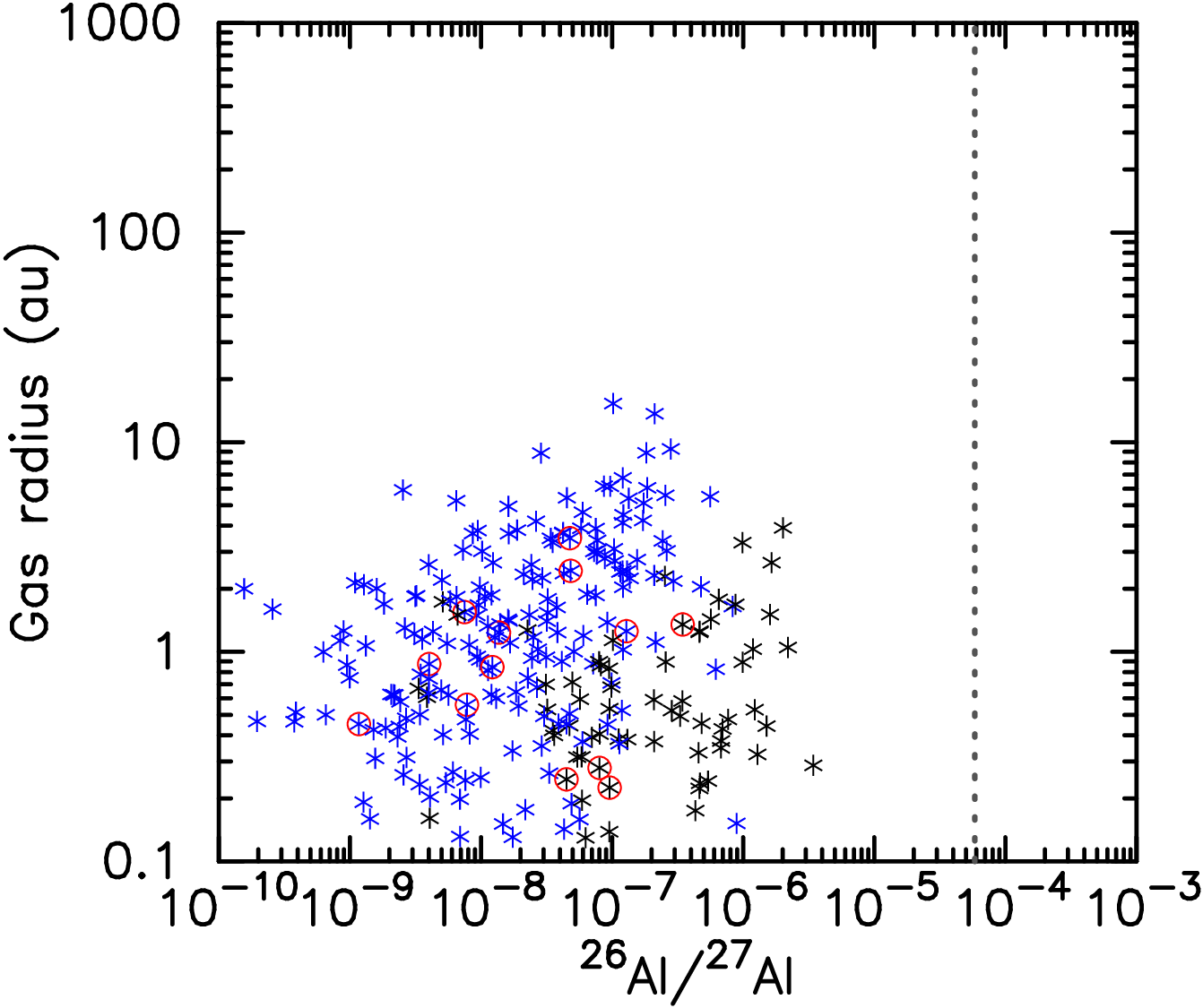}}}
\caption[bf]{Properties of discs as a function of the amount of enrichment in a moderately dense star-forming region ($\tilde{\rho} \sim 100$\,M$_\odot$\,pc$^{-3})$ where the discs all have initial radii $r_{\rm disc,i} = 100$\,au. The disc gas radii are evolved inwards due to photoevaporation (and the discs lose gas-mass), and the dust radii also follow the gas radii inwards. The dust masses remain constant. Panels (a)--(b) show the gas mass as a function of the amount of $^{26}$Al enrichment (expressed as the $^{26}$Al/$^{27}$Al ratio). Panel (a) shows enrichment from supernovae and dispersed wind ejecta and panel (b) shows enrichment from supernovae and localised wind ejecta. Panels (c)--(d) show the gas radii as a function of the amount of $^{26}$Al enrichment, where panel (c) shows enrichment from supernovae and dispersed wind ejecta and panel (d) shows enrichment from supernovae and localised wind ejecta. The vertical dotted lines in panels (a)--(d) show the measured $^{26}$Al abundance in the Solar system \cite{Thrane06}. The horizontal lines in panels (a) and (b) indicate 1\,M$_{\rm Jup}$, which we take to be the minimum mass required to form Jupiter and Saturn in our Solar system.  In all panels, black points indicate the values at 10\,Myr, whereas the blue points indicate the values much earlier, at 2.5\,Myr. Stars that are roughly Solar-mass (0.5 - 1.5\,M$_\odot$) are highlighted by a red circle.}
%Simulation 2, 10\% disc mass, 2D, distance to most massive star. 
\label{moderate_densities_inward_dust}
  \end{center}
\end{figure*}

We show the disc gas properties (radius and mass) as a function of the  amount of $^{26}$Al enrichment in Fig.~\ref{moderate_densities_inward_dust}. In this figure we do not show the results for enrichment from supernovae alone, as no discs that are in close proximity to massive stars as they explode retain any gas content. Similarly, we do not show plots for $^{60}$Fe enrichment because the gas content of discs enriched in $^{60}$Fe has been destroyed.

The lefthand panels (Fig.~\ref{moderate_densities_inward_dust-b}~and~\ref{moderate_densities_inward_dust-e}) show the enrichment versus disc mass (panel a) and enrichment versus disc radius (panel c) for dispersed winds and supernovae. As before, more discs survive at 2.5\,Myr (the blue points) than at 10\,Myr (the black points); however, the enrichment levels at 2.5\,Myr are much lower than at 10\,Myr. However, even at 10\,Myr, the enrichment levels from winds that have dispersed around the star-forming region  are always $Z_{\rm Al} < 10^{-7}$, more than two orders of magnitude lower than the measured value in the Solar system.

The enrichment levels of discs are higher  when we assume the winds are retained in localised bubbles (Fig.~\ref{moderate_densities_inward_dust-c}~and~\ref{moderate_densities_inward_dust-f}), but fewer stars' discs survive due to their closer proximity to the photoionising radiation. Even then, the enrichment levels are not close to the lower-limit or ``canonical value" of the $Z_{\rm Al}$ ratio for the Solar system \citep[$\sim 10^{-5}$,][]{Jacobsen08,Kita13}.

\subsection{Photoevaporation and viscous evolution of disc, $r_{\rm disc,i} = 10$\,au}
%2.5\,pc, 10\,au, photoevap \& viscous, LC18r3, Fig. 2}

Many theories of the evolution of protoplanetary discs posit that a significant amount of expansion due to viscosity occurs \citep[e.g.][]{Shakura73,LyndenBell74,Hartmann98,ConchaRamirez19}, although recent observations suggest that the amount of viscous expansion may be quite modest \citep[e.g.][]{Manara22}.

In similar simulations to those we present here, \citet{Parker21a} show that viscous discs that expand are depleted much faster than static discs, because the movement of the disc radius inwards preserves -- to some extent -- the surface density of the disc, whereas a reduction in mass from photoevaporation, followed by the gas radius moving back outwards, reduces the surface density of the disc. The reduction in surface density therefore makes the disc even more susceptible to photoevaporation at the next timestep.

In one set of simulations, we implemented photoevaporation, disc mass-loss and viscous evolution with an initial disc radius of $r_{\rm disc,i} = 100$\,au, but all of the discs were destroyed before any significant enrichment had occurred. Therefore, in the set of simulations we describe here, the disc radii are initialised to $r_{\rm disc,i} = 10$\,au, which was adoped by \citet{Lichtenberg16b} in their simulations.

As with the previous simulations where the dust radius is allowed to move inwards with the gas radius, the discs are either destroyed, or have such small radii when the supernovae explode, that we do not get any significant enrichment of discs from supernovae alone. Again, this means that the discs in these simulations are not enriched in $^{60}$Fe and the enrichment in $^{26}$Al comes from the stellar winds.

\begin{figure*}
  \begin{center}
\setlength{\subfigcapskip}{10pt}
\hspace*{0.2cm}\subfigure[]{\label{moderate_densities_10au-b}\rotatebox{270}{\includegraphics[scale=0.33]{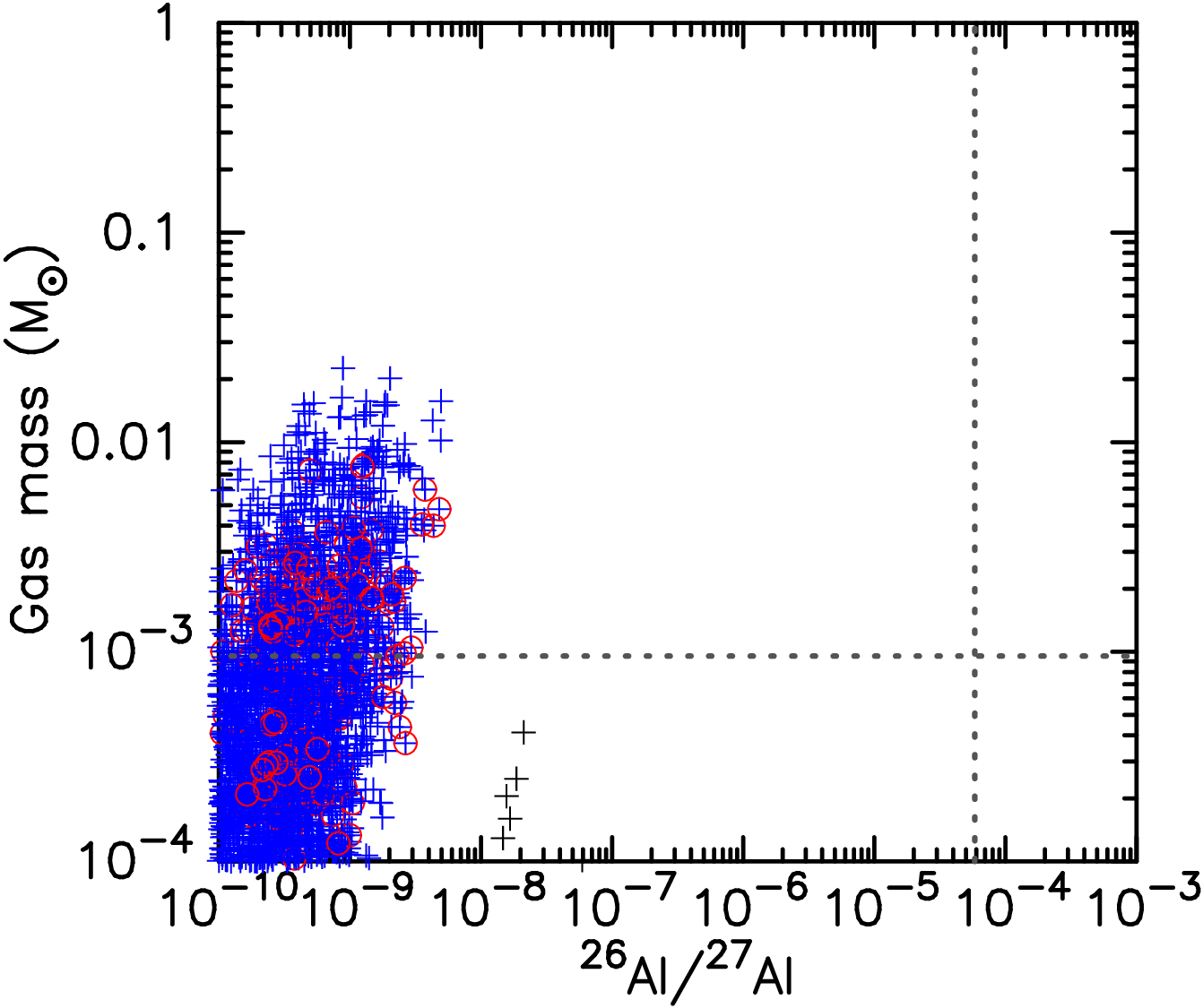}}}
\hspace*{0.2cm}\subfigure[]{\label{moderate_densities_10au-c}\rotatebox{270}{\includegraphics[scale=0.33]{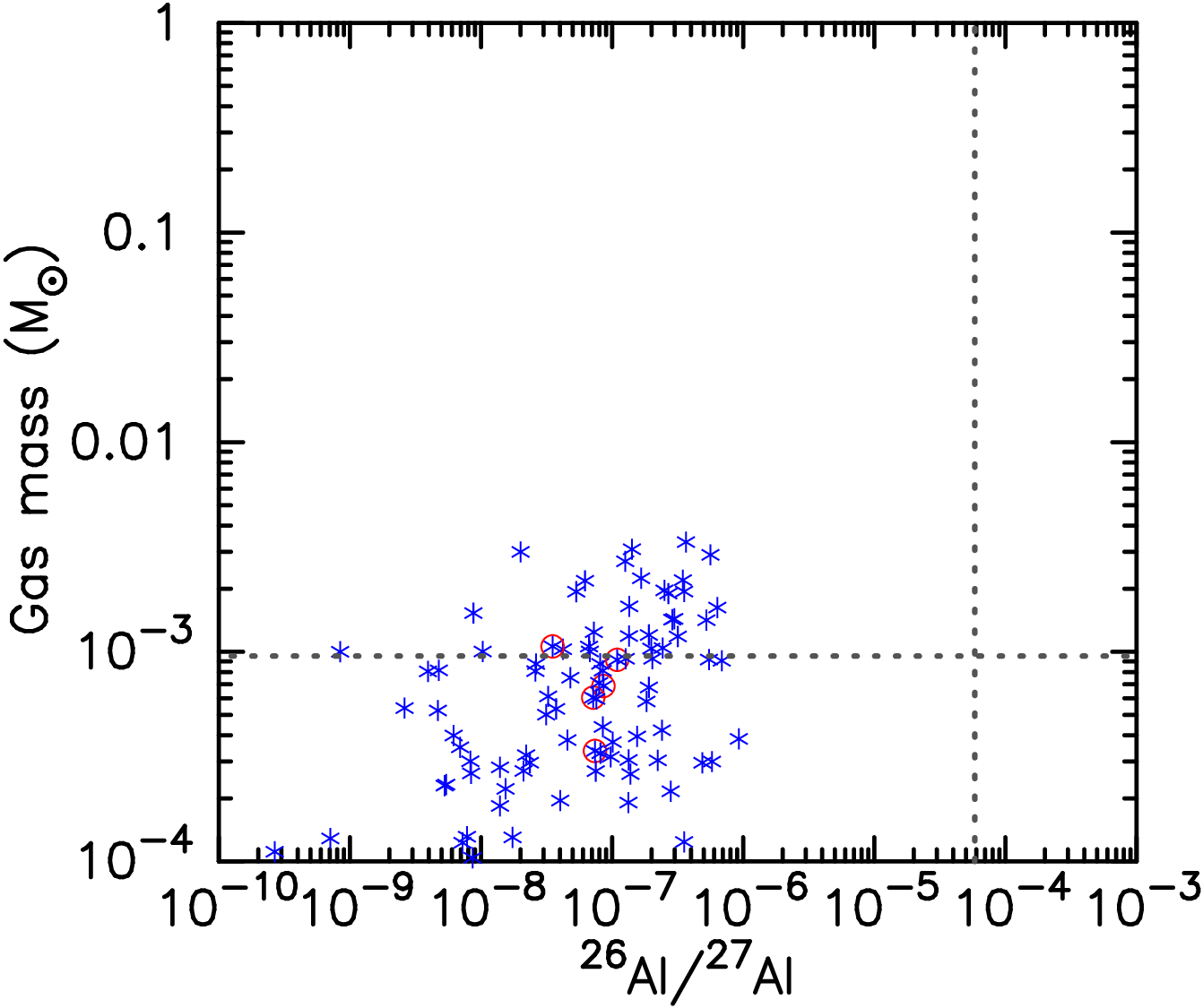}}}
\hspace*{0.2cm}\subfigure[]{\label{moderate_densities_10au-e}\rotatebox{270}{\includegraphics[scale=0.33]{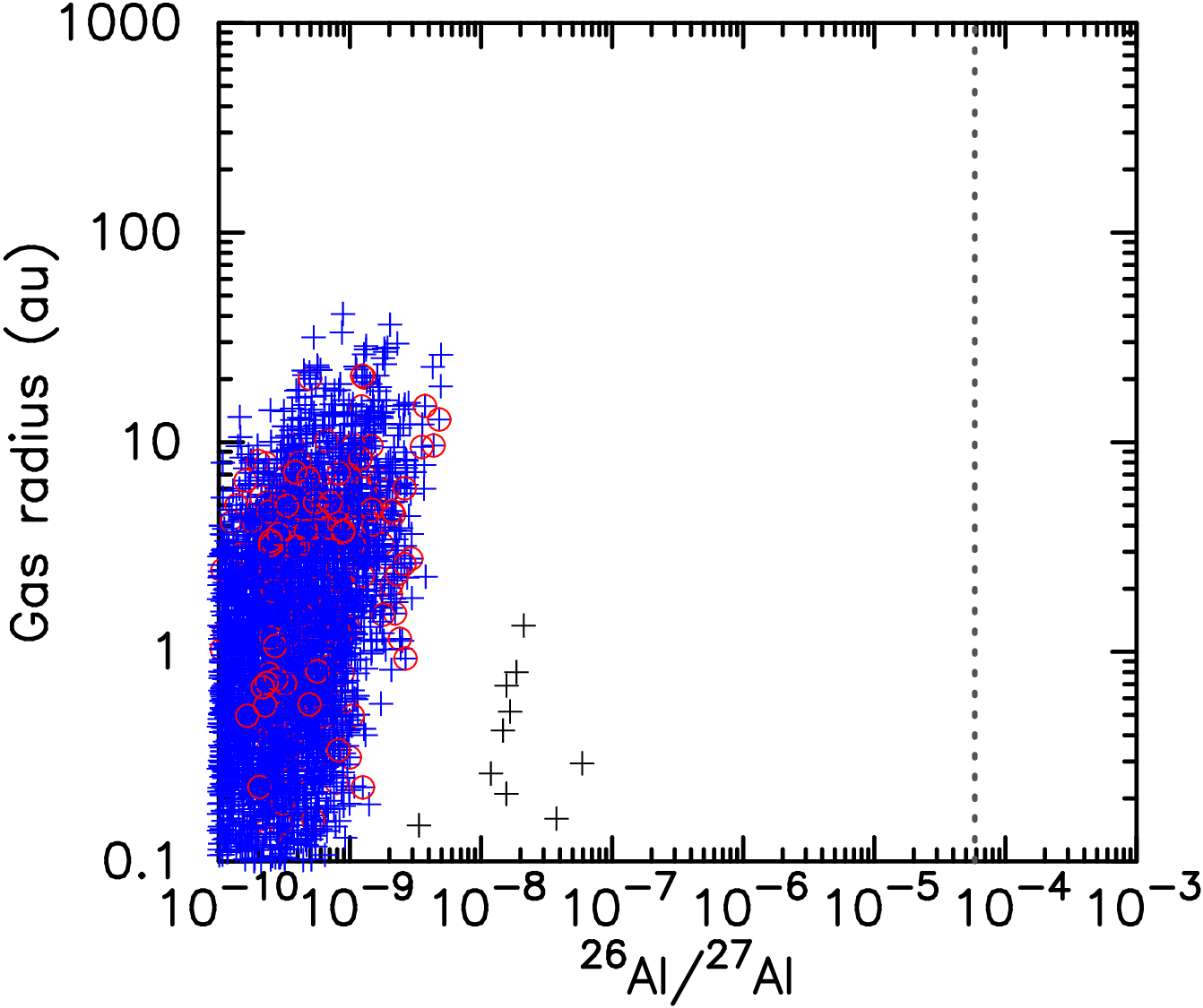}}}
\hspace*{0.2cm}\subfigure[]{\label{moderate_densities_10au-f}\rotatebox{270}{\includegraphics[scale=0.33]{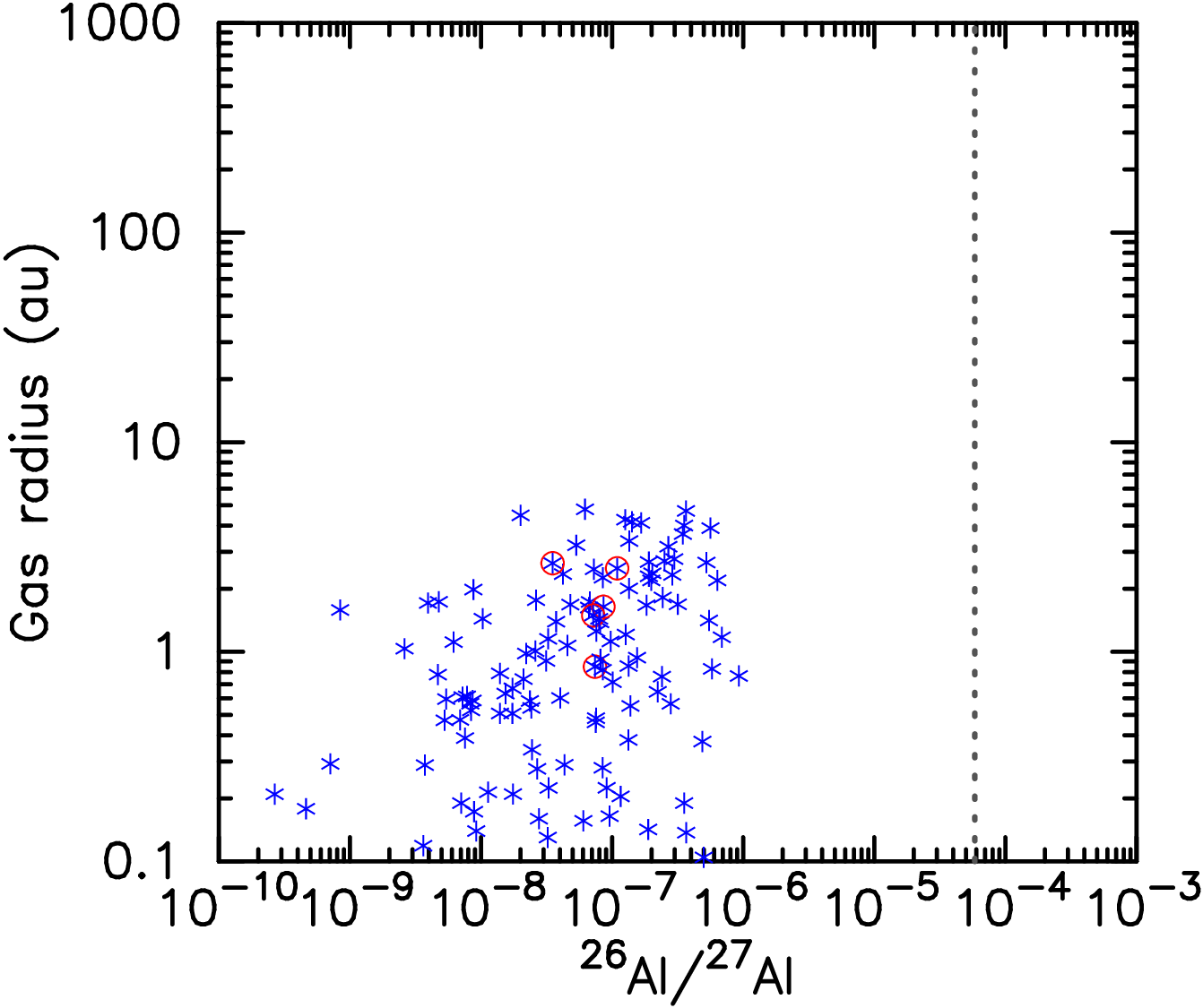}}}
\caption[bf]{Properties of discs as a function of the amount of enrichment in a moderately dense star-forming region ($\tilde{\rho} \sim 100$\,M$_\odot$\,pc$^{-3})$ where the discs all have initial radii $r_{\rm disc,i} = 10$\,au. The disc gas radii are evolved inwards due to photoevaporation (and the discs lose gas-mass), and outwards due to viscous spreading. The dust radii follow the gas radii inwards due to photoevaporation (though the dust masses remain constant) and outwards due to viscous spreading. Panels (a)--(b) show the gas mass as a function of the amount of $^{26}$Al enrichment (expressed as the $^{26}$Al/$^{27}$Al ratio). Panel (a) shows enrichment from supernovae and dispersed wind ejecta and panel (b) shows enrichment from supernovae and localised wind ejecta. Panels (c)--(d) show the gas radii as a function of the amount of $^{26}$Al enrichment, where panel (c) shows enrichment from supernovae and dispersed wind ejecta and panel (d) shows enrichment from supernovae and localised wind ejecta. The vertical dotted lines in panels (a)--(d) show the measured $^{26}$Al abundance in the Solar system \cite{Thrane06}. The horizontal lines in panels (a) and (b) indicate 1\,M$_{\rm Jup}$, which we take to be the minimum mass required to form Jupiter and Saturn in our Solar system. In all panels, black points indicate the values at 10\,Myr, whereas the blue points indicate the values much earlier, at 2.5\,Myr. Stars that are roughly Solar-mass (0.5 - 1.5\,M$_\odot$) are highlighted by a red circle.}
%Simulation 2, 10\% disc mass, 2D, distance to most massive star. 
\label{moderate_densities_10au}
  \end{center}
\end{figure*}

We show the results for the simulations where we assume the winds are dispersed in the lefthand panels (a and c) in Fig.~\ref{moderate_densities_10au}. A significant number of discs experience low levels of enrichment ($Z_{\rm Al} < 10^{-8}$) after 2.5\,Myr, but only a handful of systems that are enriched retain any gas after 10\,Myr, and even then the enrichment is $Z_{\rm Al} < 10^{-7}$, two orders of magnitude lower than in the Solar system.

If we assume that the winds from the massive stars stay in localised bubbles (Fig.~\ref{moderate_densities_10au-c}~and~\ref{moderate_densities_10au-f}), then the enrichment levels increase to $Z_{\rm Al} \leq 10^{-6}$, but this is still an order of magnitude lower than the Solar system value. Furthermore, in these models, no discs are enriched and retain gas content at 10\,Myr (note the absence of black points in Figs.~\ref{moderate_densities_10au-c}~and~\ref{moderate_densities_10au-f}).

\subsection{Stellar density}

In the results presented in the previous subsections, we adopt initial stellar densities of $\sim$100\,M$_\odot$\,pc$^{-3}$. The initial stellar density of observed star-forming regions is difficult to determine, but analysis of the current spatial \citep{Parker14b,BlaylockSquibbs22} and kinematic \citep{Schoettler20} distributions of star-forming regions can place strong constraints on the initial densities.

Recent work \citep{Parker22a} has shown that some star-forming regions may be as dense as $\sim 10^4$\,M$_\odot$\,pc$^{-3}$, whereas some very low density regions (e.g. Taurus, Cyg OB2) exhibit spatial and kinematic substructure, which means they were unlikely to have ever been more dense than they are observed today \citep{Wright14,Wright16}.

The more dense a star-forming region is, the more enrichment will occur, but conversely, there will be more destructive photoevaporation than in lower density regions. For several of our models, we also ran high density versions of the same simulation and found largely similar results; some systems experienced greater enrichment, but also lost more gas content from their discs and the two effects largely cancel each other out. Similarly, for very low-density regions ($\tilde{\rho} \sim 10$\,M$_\odot$\,pc$^{-3}$, very little photoevaporation occurs, but there is also less (usually zero) enrichment.

\subsection{Stellar evolution of massive stars}

For the majority of our initial conditions sets we calculated the yields from winds and supernovae of massive stars with stellar rotation \citep[300\,km\,s$^{-1}$; the `recommended' models from][]{Limongi18} but we also ran simulations where the massive stars have zero rotation \citep[also from][]{Limongi18}.

As detailed in \citet{Parker23a}, the effect of including the rotation of the massive stars in the models \citep[see][]{deMink13} is to prolong the lifetimes of the massive stars before they explode as supernovae, but rotation also dredges up more of the interior of the massive stars, so more SLR material is then entrained in the stellar winds.

In our simulations, when we include rotation in the massive stars we find that the levels of enrichment can be higher by an order of magnitude, but almost none of the enrichment comes from supernovae. In contrast, when the massive stars have zero rotation, they explode as supernovae earlier, but the overall enrichment is lower because not as much SLR material leaves the massive stars via their winds.

\section{Discussion}
\label{discuss}

Despite exploring a wide range of parameter space, we have been unable to enrich a protoplanetary disc with Solar system levels of $^{26}$Al and $^{60}$Fe which also retains a significant mass ($>1$\,M$_{\rm Jup}$) in gas with a radius of around 10\,au (which would facilitate the formation of gas giants like those in our Solar system). We almost match these constraints for simulations in which the dust component of the disc dynamically decouples from the gas component, and the dust radius remains static when the gas radius moves inwards.

Current observational consensus \citep[see][and references therein]{Manara22,Miotello22} is that the viscosity parameter $\alpha$ is no higher than $\sim 10^{-3}$, which would mean that the dust radius does not increase significantly due to viscous evolution. The abundance of dust substructures in observed protoplanetary disks \citep{Bae22} similarly suggests a lack of movement of at least parts of the dust mass in a substantial number of disks. Therefore, our simulations in which we keep the dust radius fixed and allow the gas radius to move inwards due to photoevaporation are likely to be the most realistic. 

However, for other scenarios for the disc physics (including viscous evolution \citep[e.g.][]{ConchaRamirez19}, or when we allow the dust radius to follow the gas radius inwards during photoevaporation, e.g.\,\,\citet{Sellek20}), the enrichment levels are not high enough for discs that contain enough mass to form the gas giants in our Solar system.

In any scenario, we do not reproduce even the measured lower limit for the $^{60}$Fe content in the Solar system, because the $^{60}$Fe is delivered exclusively by supernovae, which only explode once the majority of the discs have lost all of their gas mass.

The natural interpretation of our results is that the Solar system's enrichment in SLRs cannot be from pollution of the protosolar disc from supernovae and wind ejecta from massive stars. There are, however, several caveats to our work which we discuss below.

First, the models of mass-loss due to photoevaporation \citep[the \texttt{FRIED} grid,][]{Haworth18b} could be overestimating the amount of mass lost due to photoevaporation. Second, gas and dust leftover from star formation in the GMC (collectively referred to as extinction) could shield the protoplanetary discs from external photoevaporation \citep{Qiao23}. However, simulations of the formation of massive stars in GMCs \citep[e.g.][]{Dale12a,Dale13a} show that the massive star winds blow huge cavities in their local vicinities, which would expose protoplanetary discs to the FUV and EUV radiation.

In addition, our disc evolution and enrichment models are a post-processing analysis, so the discs are not reactive to the dynamics within the star-forming regions. Whilst attempts have been made at including realistic discs in $N$-body simulations \citep[e.g.][]{Rosotti14}, there is yet to be a self-consistent simulation that can model the dynamics of the star-forming region and the interaction between the disc and the radiation fields.

Similarly, our simulations are a purely gravitational $N$-body problem, and do not include any physics to model the formation of stars (aside from a quasi-realistic prescription to set up the spatial and kinematic distribution of the stars) or the formation of the discs.

Our results diverge from the conclusions of previous work in this area \cite[e.g.][]{Adams06,Fatuzzo08,Adams10}, who demonstrated that our Solar system could form in a relatively populous ($N_\star \sim$ 1000) star-forming region, survive photoevaporation and also obtain the measured enrichment levels in the Solar system. However, our work makes use of newer FUV-induced mass-loss models by \citet{Haworth18b}, which \citet{Parker21a} demonstrate leads to significantly more disc destruction than the older models \citep{Johnstone98,Storzer99,Scally01}. Furthermore, previous research on enrichment \citep{Adams10,Parker14a,Lichtenberg16b} invoked supernovae from massive stars ($>$25\,M$\odot$), which explode as supernovae at early ages ($<5$\,Myr) before the discs have been photoevaporated. Supernovae provide more enrichment than stellar winds, but the most recent stellar evolution models \citep{Limongi18} find that very massive stars do not explode as supernovae. The combination of these factors are responsible for the large differences beween our results and previous work in this area.

\section{Conclusions}
\label{conclude}

We present $N$-body simulations of the dynamical evolution of star-forming regions to calculate the amount of enrichment in the short-lived radioisotopes (SLRs) $^{26}$Al and $^{60}$Fe from both the stellar winds and supernovae explosions of massive stars. Simultaneously, we calculate the effect of photoionising radiation from massive stars on the gas masses and radii of the protoplanetary discs to determine whether a disc can be enriched to Solar system levels, but retain enough gas to form Jupiter and Saturn. Our conclusions are the following:

(i) The only simulations in which we can produce Solar system levels of $^{26}$Al enrichment are where the dust radius remains fixed, i.e.\,\,it does not move inwards during photoevaporation (as the gas radius does) and does not move outwards due to viscous spreading.

(ii) When we allow the dust radius to move inwards, as is thought to happen during external photoevaporation, and/or model the viscous spreading of the disc, no discs attain Solar system levels of enrichment and retain enough gas to form the giant planets.

(iii) In no simulations do discs attain the Solar system levels of $^{60}$Fe, in spite of the large uncertainty in the measured value in the Solar system. This is because $^{60}$Fe is only produced in the supernovae explosions of massive stars in our simulations, and photoevaporation has depleted the gas component of protoplanetary discs by the time of the supernovae. Indeed, in the majority of our simulations supernovae do not occur in 10\,Myr due to the high stellar rotation of the massive stars. If we switch off stellar rotation, many more discs are enriched by the supernovae, but at levels orders of magnitude lower than the Solar system values.

(iv) Taken together, our results suggest that enrichment of the Solar system via the pollution of the protoplanetary disc from winds and supernovae of massive stars in a star-forming region seems unlikely. However, we emphasise that this result is only valid for Solar-like planetary systems that host gas giant planets; enrichment in SLRs and the related geophysical processes \citep[e.g.][]{Lichtenberg19,Lichtenberg22PP7,Lichtenberg2022} could still occur in planetary systems that do not form gas giants.

(v) We also note that in the scenario in which the formation of the Solar system occurs in a series of sequential or triggered star-forming events \citep{Gounelle12}  photoevaporation from massive stars may still be problematic for forming gas giants planets.

Given the above constraints, we encourage further research into the origin of the SLRs in the Solar system, and suggest that their possible origin in Asymptotic Giant Branch stars  \citep{TrigoRodriguez09,Lugaro2018,Battino2023,Parker23b} be investigated in further detail, as this is the only scenario in which photoionising radiation from massive stars is not present.

\section*{Acknowledgements}

We thank the anonymous referee for their comments and suggestions on the original manuscript. RJP acknowledges support from the Royal Society in the form of a Dorothy Hodgkin Fellowship. TL was supported by a research grant from the Branco Weiss Foundation.

\section*{Data availability statement}

The data used to produce the plots in this paper will be shared on reasonable request to the corresponding author.

\bibliographystyle{mnras}  
\bibliography{general_ref}

\begin{thebibliography}{}
\makeatletter
\relax
\def\mn@urlcharsother{\let\do\@makeother \do\$\do\&\do\#\do\^\do\_\do\%\do\~}
\def\mn@doi{\begingroup\mn@urlcharsother \@ifnextchar [ {\mn@doi@}
  {\mn@doi@[]}}
\def\mn@doi@[#1]#2{\def\@tempa{#1}\ifx\@tempa\@empty \href
  {http://dx.doi.org/#2} {doi:#2}\else \href {http://dx.doi.org/#2} {#1}\fi
  \endgroup}
\def\mn@eprint#1#2{\mn@eprint@#1:#2::\@nil}
\def\mn@eprint@arXiv#1{\href {http://arxiv.org/abs/#1} {{\tt arXiv:#1}}}
\def\mn@eprint@dblp#1{\href {http://dblp.uni-trier.de/rec/bibtex/#1.xml}
  {dblp:#1}}
\def\mn@eprint@#1:#2:#3:#4\@nil{\def\@tempa {#1}\def\@tempb {#2}\def\@tempc
  {#3}\ifx \@tempc \@empty \let \@tempc \@tempb \let \@tempb \@tempa \fi \ifx
  \@tempb \@empty \def\@tempb {arXiv}\fi \@ifundefined
  {mn@eprint@\@tempb}{\@tempb:\@tempc}{\expandafter \expandafter \csname
  mn@eprint@\@tempb\endcsname \expandafter{\@tempc}}}

\bibitem[\protect\citeauthoryear{Adams}{Adams}{2010}]{Adams10}
Adams F.~C.,  2010, ARA\&A, 48, 47

\bibitem[\protect\citeauthoryear{Adams, Hollenbach, Laughlin  \& Gorti}{Adams
  et~al.}{2004}]{Adams04}
Adams F.~C.,  Hollenbach D.,  Laughlin G.,   Gorti U.,  2004, ApJ, 611, 360

\bibitem[\protect\citeauthoryear{Adams, Proszkow, Fatuzzo  \& Myers}{Adams
  et~al.}{2006}]{Adams06}
Adams F.~C.,  Proszkow E.~M.,  Fatuzzo M.,   Myers P.~C.,  2006, ApJ, 641, 504

\bibitem[\protect\citeauthoryear{{Alves}, {Cleeves}, {Girart}, {Zhu}, {Franco},
  {Zurlo}  \& {Caselli}}{{Alves} et~al.}{2020}]{Alves20}
{Alves} F.~O.,  {Cleeves} L.~I.,  {Girart} J.~M.,  {Zhu} Z.,  {Franco} G.
  A.~P.,  {Zurlo} A.,   {Caselli} P.,  2020, \mn@doi [\apjl]
  {10.3847/2041-8213/abc550}, \href
  {https://ui.adsabs.harvard.edu/abs/2020ApJ...904L...6A} {904, L6}

\bibitem[\protect\citeauthoryear{{Armitage}}{{Armitage}}{2000}]{Armitage00}
{Armitage} P.~J.,  2000, A\&A, 362, 968

\bibitem[\protect\citeauthoryear{{Bae}, {Isella}, {Zhu}, {Martin}, {Okuzumi}
  \& {Suriano}}{{Bae} et~al.}{2022}]{Bae22}
{Bae} J.,  {Isella} A.,  {Zhu} Z.,  {Martin} R.,  {Okuzumi} S.,   {Suriano} S.,
   2022, \mn@doi [arXiv e-prints] {10.48550/arXiv.2210.13314}, \href
  {https://ui.adsabs.harvard.edu/abs/2022arXiv221013314B} {p. arXiv:2210.13314}

\bibitem[\protect\citeauthoryear{Bate}{Bate}{2012}]{Bate12}
Bate M.~R.,  2012, MNRAS, 419, 3115

\bibitem[\protect\citeauthoryear{{Battino} et~al.,}{{Battino}
  et~al.}{2023}]{Battino2023}
{Battino} U.,  et~al., 2023, \mn@doi [\mnras] {10.1093/mnras/stad106}, \href
  {https://ui.adsabs.harvard.edu/abs/2023MNRAS.520.2436B} {520, 2436}

\bibitem[\protect\citeauthoryear{{Blaylock-Squibbs}, {Parker}, {Buckner}  \&
  {G{\"u}del}}{{Blaylock-Squibbs} et~al.}{2022}]{BlaylockSquibbs22}
{Blaylock-Squibbs} G.~A.,  {Parker} R.~J.,  {Buckner} A. S.~M.,   {G{\"u}del}
  M.,  2022, \mn@doi [\mnras] {10.1093/mnras/stab3447}, \href
  {https://ui.adsabs.harvard.edu/abs/2022MNRAS.510.2864B} {510, 2864}

\bibitem[\protect\citeauthoryear{{Boss}}{{Boss}}{2013}]{Boss13}
{Boss} A.~P.,  2013, preprint, \href
  {http://adsabs.harvard.edu/abs/2013arXiv1306.2220B} {} (\mn@eprint {arXiv}
  {1306.2220})

\bibitem[\protect\citeauthoryear{{Boss}}{{Boss}}{2017}]{Boss17}
{Boss} A.~P.,  2017, \mn@doi [\apj] {10.3847/1538-4357/aa7cf4}, \href
  {http://adsabs.harvard.edu/abs/2017ApJ...844..113B} {844, 113}

\bibitem[\protect\citeauthoryear{{Buckner} et~al.,}{{Buckner}
  et~al.}{2019}]{Buckner19}
{Buckner} A. S.~M.,  et~al., 2019, \mn@doi [\aap]
  {10.1051/0004-6361/201832936}, \href
  {https://ui.adsabs.harvard.edu/abs/2019A&A...622A.184B} {622, A184}

\bibitem[\protect\citeauthoryear{{Buser} \& {Kurucz}}{{Buser} \&
  {Kurucz}}{1992}]{Buser92}
{Buser} R.,  {Kurucz} R.~L.,  1992, \aap, \href
  {https://ui.adsabs.harvard.edu/abs/1992A&A...264..557B} {264, 557}

\bibitem[\protect\citeauthoryear{Cartwright \& Whitworth}{Cartwright \&
  Whitworth}{2004}]{Cartwright04}
Cartwright A.,  Whitworth A.~P.,  2004, MNRAS, 348, 589

\bibitem[\protect\citeauthoryear{{Chen} et~al.,}{{Chen} et~al.}{2013}]{Chen13}
{Chen} X.,  et~al., 2013, \mn@doi [ApJ] {10.1088/0004-637X/768/2/110}, \href
  {http://cdsads.u-strasbg.fr/abs/2013ApJ...768..110C} {768, 110}

\bibitem[\protect\citeauthoryear{{Concha-Ram{\'\i}rez}, {Wilhelm}, {Portegies
  Zwart}  \& {Haworth}}{{Concha-Ram{\'\i}rez} et~al.}{2019}]{ConchaRamirez19}
{Concha-Ram{\'\i}rez} F.,  {Wilhelm} M. J.~C.,  {Portegies Zwart} S.,
  {Haworth} T.~J.,  2019, \mn@doi [\mnras] {10.1093/mnras/stz2973}, \href
  {https://ui.adsabs.harvard.edu/abs/2019MNRAS.490.5678C} {490, 5678}

\bibitem[\protect\citeauthoryear{{Concha-Ram{\'\i}rez}, {Wilhelm}, {Portegies
  Zwart}, {van Terwisga}  \& {Hacar}}{{Concha-Ram{\'\i}rez}
  et~al.}{2021}]{ConchaRamirez21}
{Concha-Ram{\'\i}rez} F.,  {Wilhelm} M. J.~C.,  {Portegies Zwart} S.,  {van
  Terwisga} S.~E.,   {Hacar} A.,  2021, \mn@doi [\mnras]
  {10.1093/mnras/staa3669}, \href
  {https://ui.adsabs.harvard.edu/abs/2021MNRAS.501.1782C} {501, 1782}

\bibitem[\protect\citeauthoryear{{Cook}, {Meyer}  \&
  {Sch{\"o}nb{\"a}chler}}{{Cook} et~al.}{2021}]{Cook2021}
{Cook} D.~L.,  {Meyer} B.~S.,   {Sch{\"o}nb{\"a}chler} M.,  2021, \mn@doi
  [\apj] {10.3847/1538-4357/ac0add}, \href
  {https://ui.adsabs.harvard.edu/abs/2021ApJ...917...59C} {917, 59}

\bibitem[\protect\citeauthoryear{{Crowther}, {Schnurr}, {Hirschi}, {Yusof},
  {Parker}, {Goodwin}  \& {Kassim}}{{Crowther} et~al.}{2010}]{Crowther10}
{Crowther} P.~A.,  {Schnurr} O.,  {Hirschi} R.,  {Yusof} N.,  {Parker} R.~J.,
  {Goodwin} S.~P.,   {Kassim} H.~A.,  2010, \mn@doi [\mnras]
  {10.1111/j.1365-2966.2010.17167.x}, \href
  {https://ui.adsabs.harvard.edu/abs/2010MNRAS.408..731C} {408, 731}

\bibitem[\protect\citeauthoryear{{Dale} \& {Bonnell}}{{Dale} \&
  {Bonnell}}{2012}]{Dale12a}
{Dale} J.~E.,  {Bonnell} I.~A.,  2012, \mn@doi [MNRAS]
  {10.1111/j.1365-2966.2012.20709.x}, \href
  {http://adsabs.harvard.edu/abs/2012MNRAS.422.1352D} {422, 1352}

\bibitem[\protect\citeauthoryear{{Dale}, {Ercolano}  \& {Bonnell}}{{Dale}
  et~al.}{2013}]{Dale13a}
{Dale} J.~E.,  {Ercolano} B.,   {Bonnell} I.~A.,  2013, MNRAS, 430, 234

\bibitem[\protect\citeauthoryear{{Duch{\^e}ne} \& {Kraus}}{{Duch{\^e}ne} \&
  {Kraus}}{2013}]{Duchene13b}
{Duch{\^e}ne} G.,  {Kraus} A.,  2013, \mn@doi [ARA\&A]
  {10.1146/annurev-astro-081710-102602}, 51, 269

\bibitem[\protect\citeauthoryear{Elmegreen}{Elmegreen}{2006}]{Elmegreen06}
Elmegreen B.~G.,  2006, ApJ, 648, 572

\bibitem[\protect\citeauthoryear{{Fatuzzo} \& {Adams}}{{Fatuzzo} \&
  {Adams}}{2008}]{Fatuzzo08}
{Fatuzzo} M.,  {Adams} F.~C.,  2008, \mn@doi [ApJ] {10.1086/527469}, \href
  {http://adsabs.harvard.edu/abs/2008ApJ...675.1361F} {675, 1361}

\bibitem[\protect\citeauthoryear{{Fatuzzo} \& {Adams}}{{Fatuzzo} \&
  {Adams}}{2015}]{Fatuzzo15}
{Fatuzzo} M.,  {Adams} F.~C.,  2015, \mn@doi [\apj]
  {10.1088/0004-637X/813/1/55}, \href
  {https://ui.adsabs.harvard.edu/abs/2015ApJ...813...55F} {813, 55}

\bibitem[\protect\citeauthoryear{{Fatuzzo} \& {Adams}}{{Fatuzzo} \&
  {Adams}}{2022}]{Fatuzzo22}
{Fatuzzo} M.,  {Adams} F.~C.,  2022, \mn@doi [\apj] {10.3847/1538-4357/ac38a7},
  \href {https://ui.adsabs.harvard.edu/abs/2022ApJ...925...56F} {925, 56}

\bibitem[\protect\citeauthoryear{{Foster} et~al.,}{{Foster}
  et~al.}{2015}]{Foster15}
{Foster} J.~B.,  et~al., 2015, \mn@doi [ApJ] {10.1088/0004-637X/799/2/136},
  \href {http://adsabs.harvard.edu/abs/2015ApJ...799..136F} {799, 136}

\bibitem[\protect\citeauthoryear{{Gilmour} \& {Middleton}}{{Gilmour} \&
  {Middleton}}{2009}]{Gilmour09}
{Gilmour} J.~D.,  {Middleton} C.~A.,  2009, \mn@doi [\icarus]
  {10.1016/j.icarus.2009.03.013}, \href
  {https://ui.adsabs.harvard.edu/abs/2009Icar..201..821G} {201, 821}

\bibitem[\protect\citeauthoryear{Goodwin, Whitworth  \&
  {Ward-Thompson}}{Goodwin et~al.}{2004}]{Goodwin04c}
Goodwin S.~P.,  Whitworth A.~P.,   {Ward-Thompson} D.,  2004, A\&A, 423, 169

\bibitem[\protect\citeauthoryear{{Gounelle}}{{Gounelle}}{2015}]{Gounelle15}
{Gounelle} M.,  2015, \mn@doi [A\&A] {10.1051/0004-6361/201526174}, 582, A26

\bibitem[\protect\citeauthoryear{{Gounelle} \& {Meynet}}{{Gounelle} \&
  {Meynet}}{2012}]{Gounelle12}
{Gounelle} M.,  {Meynet} G.,  2012, \mn@doi [A\&A]
  {10.1051/0004-6361/201219031}, 545, A4

\bibitem[\protect\citeauthoryear{{Habing}}{{Habing}}{1968}]{Habing68}
{Habing} H.~J.,  1968, BAIN, \href
  {http://adsabs.harvard.edu/abs/1968BAN....19..421H} {19, 421}

\bibitem[\protect\citeauthoryear{{Hacar}, {Tafalla}, {Kauffmann}  \&
  {Kov{\'a}cs}}{{Hacar} et~al.}{2013}]{Hacar13}
{Hacar} A.,  {Tafalla} M.,  {Kauffmann} J.,   {Kov{\'a}cs} A.,  2013, A\&A,
  554, A55

\bibitem[\protect\citeauthoryear{{Haisch}, {Lada}  \& {Lada}}{{Haisch}
  et~al.}{2001}]{Haisch01}
{Haisch} Jr. K.~E.,  {Lada} E.~A.,   {Lada} C.~J.,  2001, \mn@doi [ApJL]
  {10.1086/320685}, \href {http://adsabs.harvard.edu/abs/2001ApJ...553L.153H}
  {553, L153}

\bibitem[\protect\citeauthoryear{{Hartmann}}{{Hartmann}}{2009}]{Hartmann09}
{Hartmann} L.,  2009, {Accretion Processes in Star Formation: Second Edition}

\bibitem[\protect\citeauthoryear{{Hartmann}, {Calvet}, {Gullbring}  \&
  {D'Alessio}}{{Hartmann} et~al.}{1998}]{Hartmann98}
{Hartmann} L.,  {Calvet} N.,  {Gullbring} E.,   {D'Alessio} P.,  1998, \mn@doi
  [ApJ] {10.1086/305277}, \href
  {https://ui.adsabs.harvard.edu/abs/1998ApJ...495..385H} {495, 385}

\bibitem[\protect\citeauthoryear{{Haworth} \& {Clarke}}{{Haworth} \&
  {Clarke}}{2019}]{Haworth19}
{Haworth} T.~J.,  {Clarke} C.~J.,  2019, \mn@doi [\mnras]
  {10.1093/mnras/stz706}, \href
  {https://ui.adsabs.harvard.edu/abs/2019MNRAS.485.3895H} {485, 3895}

\bibitem[\protect\citeauthoryear{{Haworth}, {Facchini}, {Clarke}  \&
  {Mohanty}}{{Haworth} et~al.}{2018a}]{Haworth18a}
{Haworth} T.~J.,  {Facchini} S.,  {Clarke} C.~J.,   {Mohanty} S.,  2018a,
  \mn@doi [\mnras] {10.1093/mnras/sty168}, \href
  {http://adsabs.harvard.edu/abs/2018MNRAS.475.5460H} {475, 5460}

\bibitem[\protect\citeauthoryear{{Haworth}, {Clarke}, {Rahman}, {Winter}  \&
  {Facchini}}{{Haworth} et~al.}{2018b}]{Haworth18b}
{Haworth} T.~J.,  {Clarke} C.~J.,  {Rahman} W.,  {Winter} A.~J.,   {Facchini}
  S.,  2018b, \mn@doi [\mnras] {10.1093/mnras/sty2323}, \href
  {http://adsabs.harvard.edu/abs/2018MNRAS.481..452H} {481, 452}

\bibitem[\protect\citeauthoryear{{Jacobsen}, {Yin}, {Moynier}, {Amelin},
  {Krot}, {Nagashima}, {Hutcheon}  \& {Palme}}{{Jacobsen}
  et~al.}{2008}]{Jacobsen08}
{Jacobsen} B.,  {Yin} Q.-z.,  {Moynier} F.,  {Amelin} Y.,  {Krot} A.~N.,
  {Nagashima} K.,  {Hutcheon} I.~D.,   {Palme} H.,  2008, \mn@doi [Earth and
  Planetary Science Letters] {10.1016/j.epsl.2008.05.003}, \href
  {https://ui.adsabs.harvard.edu/abs/2008E&PSL.272..353J} {272, 353}

\bibitem[\protect\citeauthoryear{{Johnstone}, {Hollenbach}  \&
  {Bally}}{{Johnstone} et~al.}{1998}]{Johnstone98}
{Johnstone} D.,  {Hollenbach} D.,   {Bally} J.,  1998, \mn@doi [\apj]
  {10.1086/305658}, \href
  {https://ui.adsabs.harvard.edu/abs/1998ApJ...499..758J} {499, 758}

\bibitem[\protect\citeauthoryear{Kirk \& Myers}{Kirk \& Myers}{2011}]{Kirk10}
Kirk H.,  Myers P.~C.,  2011, ApJ, 727, 64

\bibitem[\protect\citeauthoryear{{Kita} et~al.,}{{Kita} et~al.}{2013}]{Kita13}
{Kita} N.~T.,  et~al., 2013, \mn@doi [Meteoritics and Planetary Science]
  {10.1111/maps.12141}, \href
  {http://adsabs.harvard.edu/abs/2013M%26PS...48.1383K} {48, 1383}

\bibitem[\protect\citeauthoryear{Lada \& Lada}{Lada \& Lada}{2003}]{Lada03}
Lada C.~J.,  Lada E.~A.,  2003, ARA\&A, 41, 57

\bibitem[\protect\citeauthoryear{Larson}{Larson}{1981}]{Larson81}
Larson R.~B.,  1981, MNRAS, 194, 809

\bibitem[\protect\citeauthoryear{{Lehmann}, {Vitrichenko}, {Bychkov},
  {Bychkova}  \& {Klochkova}}{{Lehmann} et~al.}{2010}]{Lehmann10}
{Lehmann} H.,  {Vitrichenko} E.,  {Bychkov} V.,  {Bychkova} L.,   {Klochkova}
  V.,  2010, \mn@doi [\aap] {10.1051/0004-6361/201013992}, \href
  {https://ui.adsabs.harvard.edu/abs/2010A&A...514A..34L} {514, A34}

\bibitem[\protect\citeauthoryear{{Lichtenberg} \& {Clement}}{{Lichtenberg} \&
  {Clement}}{2022}]{Lichtenberg2022}
{Lichtenberg} T.,  {Clement} M.~S.,  2022, \mn@doi [\apjl]
  {10.3847/2041-8213/ac9521}, \href
  {https://ui.adsabs.harvard.edu/abs/2022ApJ...938L...3L} {938, L3}

\bibitem[\protect\citeauthoryear{Lichtenberg, Parker  \& Meyer}{Lichtenberg
  et~al.}{2016}]{Lichtenberg16b}
Lichtenberg T.,  Parker R.~J.,   Meyer M.~R.,  2016, MNRAS, 462, 3979

\bibitem[\protect\citeauthoryear{{Lichtenberg}, {Golabek}, {Burn}, {Meyer},
  {Alibert}, {Gerya}  \& {Mordasini}}{{Lichtenberg}
  et~al.}{2019}]{Lichtenberg19}
{Lichtenberg} T.,  {Golabek} G.~J.,  {Burn} R.,  {Meyer} M.~R.,  {Alibert} Y.,
  {Gerya} T.~V.,   {Mordasini} C.,  2019, \mn@doi [Nature Astronomy]
  {10.1038/s41550-018-0688-5}, \href
  {https://ui.adsabs.harvard.edu/abs/2019NatAs...3..307L} {3, 307}

\bibitem[\protect\citeauthoryear{{Lichtenberg}, {Schaefer}, {Nakajima}  \&
  {Fischer}}{{Lichtenberg} et~al.}{2022}]{Lichtenberg22PP7}
{Lichtenberg} T.,  {Schaefer} L.~K.,  {Nakajima} M.,   {Fischer} R.~A.,  2022,
  \mn@doi [arXiv e-prints] {10.48550/arXiv.2203.10023}, \href
  {https://ui.adsabs.harvard.edu/abs/2022arXiv220310023L} {p. arXiv:2203.10023}

\bibitem[\protect\citeauthoryear{{Limongi} \& {Chieffi}}{{Limongi} \&
  {Chieffi}}{2018}]{Limongi18}
{Limongi} M.,  {Chieffi} A.,  2018, \mn@doi [\apjs] {10.3847/1538-4365/aacb24},
  \href {https://ui.adsabs.harvard.edu/abs/2018ApJS..237...13L} {237, 13}

\bibitem[\protect\citeauthoryear{{Lodders}}{{Lodders}}{2003}]{Lodders03}
{Lodders} K.,  2003, \mn@doi [\apj] {10.1086/375492}, \href
  {https://ui.adsabs.harvard.edu/abs/2003ApJ...591.1220L} {591, 1220}

\bibitem[\protect\citeauthoryear{{Lugaro}, {Ott}  \& {Kereszturi}}{{Lugaro}
  et~al.}{2018}]{Lugaro2018}
{Lugaro} M.,  {Ott} U.,   {Kereszturi} {\'A}.,  2018, \mn@doi [Progress in
  Particle and Nuclear Physics] {10.1016/j.ppnp.2018.05.002}, \href
  {https://ui.adsabs.harvard.edu/abs/2018PrPNP.102....1L} {102, 1}

\bibitem[\protect\citeauthoryear{{Luhman}}{{Luhman}}{2004}]{Luhman04a}
{Luhman} K.~L.,  2004, \mn@doi [\apj] {10.1086/381146}, \href
  {https://ui.adsabs.harvard.edu/abs/2004ApJ...602..816L} {602, 816}

\bibitem[\protect\citeauthoryear{{Luhman}, {Stauffer}, {Muench}, {Rieke},
  {Lada}, {Bouvier}  \& {Lada}}{{Luhman} et~al.}{2003}]{Luhman03b}
{Luhman} K.~L.,  {Stauffer} J.~R.,  {Muench} A.~A.,  {Rieke} G.~H.,  {Lada}
  E.~A.,  {Bouvier} J.,   {Lada} C.~J.,  2003, \mn@doi [ApJ] {10.1086/376594},
  \href {http://adsabs.harvard.edu/abs/2003ApJ...593.1093L} {593, 1093}

\bibitem[\protect\citeauthoryear{{Lynden-Bell} \& {Pringle}}{{Lynden-Bell} \&
  {Pringle}}{1974}]{LyndenBell74}
{Lynden-Bell} D.,  {Pringle} J.~E.,  1974, \mn@doi [MNRAS]
  {10.1093/mnras/168.3.603}, \href
  {https://ui.adsabs.harvard.edu/abs/1974MNRAS.168..603L} {168, 603}

\bibitem[\protect\citeauthoryear{{Manara}, {Ansdell}, {Rosotti}, {Hughes},
  {Armitage}, {Lodato}  \& {Williams}}{{Manara} et~al.}{2022}]{Manara22}
{Manara} C.~F.,  {Ansdell} M.,  {Rosotti} G.~P.,  {Hughes} A.~M.,  {Armitage}
  P.~J.,  {Lodato} G.,   {Williams} J.~P.,  2022, \mn@doi [arXiv e-prints]
  {10.48550/arXiv.2203.09930}, \href
  {https://ui.adsabs.harvard.edu/abs/2022arXiv220309930M} {p. arXiv:2203.09930}

\bibitem[\protect\citeauthoryear{Maschberger}{Maschberger}{2013}]{Maschberger13}
Maschberger T.,  2013, MNRAS, 429, 1725

\bibitem[\protect\citeauthoryear{{Miotello}, {Kamp}, {Birnstiel}, {Cleeves}  \&
  {Kataoka}}{{Miotello} et~al.}{2022}]{Miotello22}
{Miotello} A.,  {Kamp} I.,  {Birnstiel} T.,  {Cleeves} L.~I.,   {Kataoka} A.,
  2022, \mn@doi [arXiv e-prints] {10.48550/arXiv.2203.09818}, \href
  {https://ui.adsabs.harvard.edu/abs/2022arXiv220309818M} {p. arXiv:2203.09818}

\bibitem[\protect\citeauthoryear{{Mishra}, {Marhas}  \& {Sameer}}{{Mishra}
  et~al.}{2016}]{Mishra16}
{Mishra} R.~K.,  {Marhas} K.~K.,   {Sameer} 2016, \mn@doi [Earth and Planetary
  Science Letters] {10.1016/j.epsl.2015.12.007}, \href
  {https://ui.adsabs.harvard.edu/abs/2016E&PSL.436...71M} {436, 71}

\bibitem[\protect\citeauthoryear{{Najita} \& {Kenyon}}{{Najita} \&
  {Kenyon}}{2014}]{Najita2014}
{Najita} J.~R.,  {Kenyon} S.~J.,  2014, \mn@doi [\mnras]
  {10.1093/mnras/stu1994}, \href
  {https://ui.adsabs.harvard.edu/abs/2014MNRAS.445.3315N} {445, 3315}

\bibitem[\protect\citeauthoryear{Nicholson \& Parker}{Nicholson \&
  Parker}{2017}]{Nicholson17}
Nicholson R.~B.,  Parker R.~J.,  2017, MNRAS, 464, 4318

\bibitem[\protect\citeauthoryear{{Nicholson}, {Parker}, {Church}, {Davies},
  {Fearon}  \& {Walton}}{{Nicholson} et~al.}{2019}]{Nicholson19a}
{Nicholson} R.~B.,  {Parker} R.~J.,  {Church} R.~P.,  {Davies} M.~B.,  {Fearon}
  N.~M.,   {Walton} S. R.~J.,  2019, \mn@doi [\mnras] {10.1093/mnras/stz606},
  \href {https://ui.adsabs.harvard.edu/abs/2019MNRAS.485.4893N} {485, 4893}

\bibitem[\protect\citeauthoryear{{Ouellette}, {Desch}  \& {Hester}}{{Ouellette}
  et~al.}{2007}]{Ouellette07}
{Ouellette} N.,  {Desch} S.~J.,   {Hester} J.~J.,  2007, \mn@doi [ApJ]
  {10.1086/518102}, \href {http://adsabs.harvard.edu/abs/2007ApJ...662.1268O}
  {662, 1268}

\bibitem[\protect\citeauthoryear{{Ouellette}, {Desch}  \& {Hester}}{{Ouellette}
  et~al.}{2010}]{Ouellette10}
{Ouellette} N.,  {Desch} S.~J.,   {Hester} J.~J.,  2010, \mn@doi [ApJ]
  {10.1088/0004-637X/711/2/597}, \href
  {https://ui.adsabs.harvard.edu/abs/2010ApJ...711..597O} {711, 597}

\bibitem[\protect\citeauthoryear{Parker \& Dale}{Parker \&
  Dale}{2015}]{Parker15c}
Parker R.~J.,  Dale J.~E.,  2015, MNRAS, 451, 3664

\bibitem[\protect\citeauthoryear{Parker \& Goodwin}{Parker \&
  Goodwin}{2007}]{Parker07}
Parker R.~J.,  Goodwin S.~P.,  2007, MNRAS, 380, 1271

\bibitem[\protect\citeauthoryear{{Parker} \& {Schoettler}}{{Parker} \&
  {Schoettler}}{2022}]{Parker22a}
{Parker} R.~J.,  {Schoettler} C.,  2022, \mn@doi [\mnras]
  {10.1093/mnras/stab3460}, \href
  {https://ui.adsabs.harvard.edu/abs/2022MNRAS.510.1136P} {510, 1136}

\bibitem[\protect\citeauthoryear{{Parker} \& {Schoettler}}{{Parker} \&
  {Schoettler}}{2023}]{Parker23b}
{Parker} R.~J.,  {Schoettler} C.,  2023, \mn@doi [\apjl]
  {10.3847/2041-8213/ace24a}, \href
  {https://ui.adsabs.harvard.edu/abs/2023ApJ...952L..16P} {952, L16}

\bibitem[\protect\citeauthoryear{Parker, Church, Davies  \& Meyer}{Parker
  et~al.}{2014a}]{Parker14a}
Parker R.~J.,  Church R.~P.,  Davies M.~B.,   Meyer M.~R.,  2014a, MNRAS, 437,
  946

\bibitem[\protect\citeauthoryear{Parker, Wright, Goodwin  \& Meyer}{Parker
  et~al.}{2014b}]{Parker14b}
Parker R.~J.,  Wright N.~J.,  Goodwin S.~P.,   Meyer M.~R.,  2014b, MNRAS, 438,
  620

\bibitem[\protect\citeauthoryear{{Parker}, {Nicholson}  \& {Alcock}}{{Parker}
  et~al.}{2021}]{Parker21a}
{Parker} R.~J.,  {Nicholson} R.~B.,   {Alcock} H.~L.,  2021, \mn@doi [\mnras]
  {10.1093/mnras/stab054}, \href
  {https://ui.adsabs.harvard.edu/abs/2021MNRAS.tmp..106P} {502, 2665}

\bibitem[\protect\citeauthoryear{{Parker}, {Lichtenberg}, {Patel}, {Polius}  \&
  {Ridsdill-Smith}}{{Parker} et~al.}{2023}]{Parker23a}
{Parker} R.~J.,  {Lichtenberg} T.,  {Patel} M.,  {Polius} C. K.~M.,
  {Ridsdill-Smith} M.,  2023, \mn@doi [\mnras] {10.1093/mnras/stad871}, \href
  {https://ui.adsabs.harvard.edu/abs/2023MNRAS.521.4838P} {521, 4838}

\bibitem[\protect\citeauthoryear{{Peretto}, {Andr{\'e}}  \&
  {Belloche}}{{Peretto} et~al.}{2006}]{Peretto06}
{Peretto} N.,  {Andr{\'e}} P.,   {Belloche} A.,  2006, \mn@doi [A\&A]
  {10.1051/0004-6361:20053324}, \href
  {http://adsabs.harvard.edu/abs/2006A\%26A...445..979P} {445, 979}

\bibitem[\protect\citeauthoryear{{Pineda} et~al.,}{{Pineda}
  et~al.}{2015}]{Pineda15}
{Pineda} J.~E.,  et~al., 2015, \mn@doi [Nature] {10.1038/nature14166}, \href
  {http://adsabs.harvard.edu/abs/2015Natur.518..213P} {518, 213}

\bibitem[\protect\citeauthoryear{{Portegies Zwart}}{{Portegies
  Zwart}}{2019}]{Zwart19}
{Portegies Zwart} S.,  2019, \mn@doi [\aap] {10.1051/0004-6361/201833974},
  \href {https://ui.adsabs.harvard.edu/abs/2019A&A...622A..69P} {622, A69}

\bibitem[\protect\citeauthoryear{{Portegies Zwart}, Makino, McMillan  \&
  Hut}{{Portegies Zwart} et~al.}{1999}]{Zwart99}
{Portegies Zwart} S.~F.,  Makino J.,  McMillan S. L.~W.,   Hut P.,  1999, A\&A,
  348, 117

\bibitem[\protect\citeauthoryear{{Portegies Zwart}, McMillan, Hut  \&
  Makino}{{Portegies Zwart} et~al.}{2001}]{Zwart01}
{Portegies Zwart} S.~F.,  McMillan S. L.~W.,  Hut P.,   Makino J.,  2001,
  MNRAS, 321, 199

\bibitem[\protect\citeauthoryear{{Pringle}}{{Pringle}}{1981}]{Pringle81}
{Pringle} J.~E.,  1981, \mn@doi [\araa] {10.1146/annurev.aa.19.090181.001033},
  \href {https://ui.adsabs.harvard.edu/abs/1981ARA&A..19..137P} {19, 137}

\bibitem[\protect\citeauthoryear{{Qiao}, {Coleman}  \& {Haworth}}{{Qiao}
  et~al.}{2023}]{Qiao23}
{Qiao} L.,  {Coleman} G. A.~L.,   {Haworth} T.~J.,  2023, \mn@doi [\mnras]
  {10.1093/mnras/stad944}, \href
  {https://ui.adsabs.harvard.edu/abs/2023MNRAS.522.1939Q} {522, 1939}

\bibitem[\protect\citeauthoryear{Raghavan et~al.,}{Raghavan
  et~al.}{2010}]{Raghavan10}
Raghavan D.,  et~al., 2010, ApJSS, 190, 1

\bibitem[\protect\citeauthoryear{{Richert}, {Getman}, {Feigelson}, {Kuhn},
  {Broos}, {Povich}, {Bate}  \& {Garmire}}{{Richert} et~al.}{2018}]{Richert18}
{Richert} A.~J.~W.,  {Getman} K.~V.,  {Feigelson} E.~D.,  {Kuhn} M.~A.,
  {Broos} P.~S.,  {Povich} M.~S.,  {Bate} M.~R.,   {Garmire} G.~P.,  2018,
  \mn@doi [\mnras] {10.1093/mnras/sty949}, \href
  {http://adsabs.harvard.edu/abs/2018MNRAS.477.5191R} {477, 5191}

\bibitem[\protect\citeauthoryear{{Rosotti}, {Dale}, {de Juan Ovelar}, {Hubber},
  {Kruijssen}, {Ercolano}  \& {Walch}}{{Rosotti} et~al.}{2014}]{Rosotti14}
{Rosotti} G.~P.,  {Dale} J.~E.,  {de Juan Ovelar} M.,  {Hubber} D.~A.,
  {Kruijssen} J.~M.~D.,  {Ercolano} B.,   {Walch} S.,  2014, \mn@doi [MNRAS]
  {10.1093/mnras/stu679}, \href
  {http://adsabs.harvard.edu/abs/2014MNRAS.441.2094R} {441, 2094}

\bibitem[\protect\citeauthoryear{Salpeter}{Salpeter}{1955}]{Salpeter55}
Salpeter E.~E.,  1955, ApJ, 121, 161

\bibitem[\protect\citeauthoryear{S{\'a}nchez \& Alfaro}{S{\'a}nchez \&
  Alfaro}{2009}]{Sanchez09}
S{\'a}nchez N.,  Alfaro E.~J.,  2009, ApJ, 696, 2086

\bibitem[\protect\citeauthoryear{Scally \& Clarke}{Scally \&
  Clarke}{2001}]{Scally01}
Scally A.,  Clarke C.,  2001, MNRAS, 325, 449

\bibitem[\protect\citeauthoryear{Scally \& Clarke}{Scally \&
  Clarke}{2002}]{Scally02}
Scally A.,  Clarke C.,  2002, MNRAS, 334, 156

\bibitem[\protect\citeauthoryear{{Schaller}, {Schaerer}, {Meynet}  \&
  {Maeder}}{{Schaller} et~al.}{1992}]{Schaller92}
{Schaller} G.,  {Schaerer} D.,  {Meynet} G.,   {Maeder} A.,  1992, \aaps, \href
  {https://ui.adsabs.harvard.edu/abs/1992A&AS...96..269S} {96, 269}

\bibitem[\protect\citeauthoryear{{Schoettler}, {de Bruijne}, {Vaher}  \&
  {Parker}}{{Schoettler} et~al.}{2020}]{Schoettler20}
{Schoettler} C.,  {de Bruijne} J.,  {Vaher} E.,   {Parker} R.~J.,  2020,
  \mn@doi [\mnras] {10.1093/mnras/staa1228}, \href
  {https://ui.adsabs.harvard.edu/abs/2020MNRAS.tmp.1474S} {495, 3104}

\bibitem[\protect\citeauthoryear{{Segura-Cox} et~al.,}{{Segura-Cox}
  et~al.}{2020}]{SeguraCox20}
{Segura-Cox} D.~M.,  et~al., 2020, \mn@doi [\nat] {10.1038/s41586-020-2779-6},
  \href {https://ui.adsabs.harvard.edu/abs/2020Natur.586..228S} {586, 228}

\bibitem[\protect\citeauthoryear{{Sellek}, {Booth}  \& {Clarke}}{{Sellek}
  et~al.}{2020}]{Sellek20}
{Sellek} A.~D.,  {Booth} R.~A.,   {Clarke} C.~J.,  2020, \mn@doi [\mnras]
  {10.1093/mnras/stz3528}, \href
  {https://ui.adsabs.harvard.edu/abs/2020MNRAS.492.1279S} {492, 1279}

\bibitem[\protect\citeauthoryear{{Shakura} \& {Sunyaev}}{{Shakura} \&
  {Sunyaev}}{1973}]{Shakura73}
{Shakura} N.~I.,  {Sunyaev} R.~A.,  1973, A\&A, \href
  {https://ui.adsabs.harvard.edu/abs/1973A&A....24..337S} {500, 33}

\bibitem[\protect\citeauthoryear{{Sternberg}, {Hoffmann}  \&
  {Pauldrach}}{{Sternberg} et~al.}{2003}]{Sternberg03}
{Sternberg} A.,  {Hoffmann} T.~L.,   {Pauldrach} A.~W.~A.,  2003, \mn@doi
  [\apj] {10.1086/379506}, \href
  {http://adsabs.harvard.edu/abs/2003ApJ...599.1333S} {599, 1333}

\bibitem[\protect\citeauthoryear{{St{\"o}rzer} \& {Hollenbach}}{{St{\"o}rzer}
  \& {Hollenbach}}{1999}]{Storzer99}
{St{\"o}rzer} H.,  {Hollenbach} D.,  1999, \mn@doi [\apj] {10.1086/307055},
  \href {http://adsabs.harvard.edu/abs/1999ApJ...515..669S} {515, 669}

\bibitem[\protect\citeauthoryear{{Tang} \& {Dauphas}}{{Tang} \&
  {Dauphas}}{2012}]{Tang12}
{Tang} H.,  {Dauphas} N.,  2012, \mn@doi [Earth and Planetary Science Letters]
  {10.1016/j.epsl.2012.10.011}, 359, 248

\bibitem[\protect\citeauthoryear{{Thrane}, {Bizzarro}  \& {Baker}}{{Thrane}
  et~al.}{2006}]{Thrane06}
{Thrane} K.,  {Bizzarro} M.,   {Baker} J.~A.,  2006, \mn@doi [ApJL]
  {10.1086/506910}, \href {http://adsabs.harvard.edu/abs/2006ApJ...646L.159T}
  {646, L159}

\bibitem[\protect\citeauthoryear{{Trappitsch} et~al.,}{{Trappitsch}
  et~al.}{2018}]{Trappitsch2018}
{Trappitsch} R.,  et~al., 2018, \mn@doi [\apjl] {10.3847/2041-8213/aabba9},
  \href {https://ui.adsabs.harvard.edu/abs/2018ApJ...857L..15T} {857, L15}

\bibitem[\protect\citeauthoryear{{Trigo-Rodr{\'\i}guez},
  {Garc{\'\i}a-Hern{\'a}ndez}, {Lugaro}, {Karakas}, {van Raai}, {Garc{\'\i}a
  Lario}  \& {Manchado}}{{Trigo-Rodr{\'\i}guez}
  et~al.}{2009}]{TrigoRodriguez09}
{Trigo-Rodr{\'\i}guez} J.~M.,  {Garc{\'\i}a-Hern{\'a}ndez} D.~A.,  {Lugaro} M.,
   {Karakas} A.~I.,  {van Raai} M.,  {Garc{\'\i}a Lario} P.,   {Manchado} A.,
  2009, \mn@doi [Meteoritics and Planetary Science]
  {10.1111/j.1945-5100.2009.tb00758.x}, \href
  {https://ui.adsabs.harvard.edu/abs/2009M&PS...44..627T} {44, 627}

\bibitem[\protect\citeauthoryear{{Vacca}, {Garmany}  \& {Shull}}{{Vacca}
  et~al.}{1996}]{Vacca96}
{Vacca} W.~D.,  {Garmany} C.~D.,   {Shull} J.~M.,  1996, \mn@doi [\apj]
  {10.1086/177020}, \href {http://adsabs.harvard.edu/abs/1996ApJ...460..914V}
  {460, 914}

\bibitem[\protect\citeauthoryear{{Ward-Duong} et~al.,}{{Ward-Duong}
  et~al.}{2015}]{Ward-Duong15}
{Ward-Duong} K.,  et~al., 2015, MNRAS, \href
  {http://adsabs.harvard.edu/abs/2015MNRAS.449.2618W} {449, 2618}

\bibitem[\protect\citeauthoryear{Weidner \& Kroupa}{Weidner \&
  Kroupa}{2006}]{Weidner06}
Weidner C.,  Kroupa P.,  2006, MNRAS, 365, 1333

\bibitem[\protect\citeauthoryear{{Winter}, {Clarke}, {Rosotti}, {Ih},
  {Facchini}  \& {Haworth}}{{Winter} et~al.}{2018}]{Winter18b}
{Winter} A.~J.,  {Clarke} C.~J.,  {Rosotti} G.,  {Ih} J.,  {Facchini} S.,
  {Haworth} T.~J.,  2018, \mn@doi [\mnras] {10.1093/mnras/sty984}, \href
  {http://adsabs.harvard.edu/abs/2018MNRAS.478.2700W} {478, 2700}

\bibitem[\protect\citeauthoryear{Wright, Parker, Goodwin  \& Drake}{Wright
  et~al.}{2014}]{Wright14}
Wright N.~J.,  Parker R.~J.,  Goodwin S.~P.,   Drake J.~J.,  2014, MNRAS, 438,
  639

\bibitem[\protect\citeauthoryear{{Wright}, {Bouy}, {Drew}, {Sarro}, {Bertin},
  {Cuillandre}  \& {Barrado}}{{Wright} et~al.}{2016}]{Wright16}
{Wright} N.~J.,  {Bouy} H.,  {Drew} J.~E.,  {Sarro} L.~M.,  {Bertin} E.,
  {Cuillandre} J.-C.,   {Barrado} D.,  2016, \mn@doi [MNRAS]
  {10.1093/mnras/stw1148}, \href
  {http://adsabs.harvard.edu/abs/2016MNRAS.460.2593W} {460, 2593}

\bibitem[\protect\citeauthoryear{{Young}}{{Young}}{2014}]{Young2014}
{Young} E.~D.,  2014, \mn@doi [Earth and Planetary Science Letters]
  {10.1016/j.epsl.2014.02.014}, \href
  {https://ui.adsabs.harvard.edu/abs/2014E&PSL.392...16Y} {392, 16}

\bibitem[\protect\citeauthoryear{{de Mink}, {Langer}, {Izzard}, {Sana}  \& {de
  Koter}}{{de Mink} et~al.}{2013}]{deMink13}
{de Mink} S.~E.,  {Langer} N.,  {Izzard} R.~G.,  {Sana} H.,   {de Koter} A.,
  2013, \mn@doi [\apj] {10.1088/0004-637X/764/2/166}, \href
  {https://ui.adsabs.harvard.edu/abs/2013ApJ...764..166D} {764, 166}

\makeatother
\end{thebibliography}

%\appendix

\label{lastpage}

\end{document}